\documentclass[referee]{aa}
\usepackage{graphicx}
\usepackage{txfonts}
\usepackage[colorlinks=true,
			linkcolor=blue,
			citecolor=blue, 
			filecolor=blue, 
			urlcolor=blue]{hyperref}
\usepackage[switch]{lineno}
\usepackage{txfonts}
\usepackage{natbib}
\usepackage{graphicx, xspace}
\usepackage{multirow}
\usepackage{ amssymb }
\usepackage{subfigure}
\usepackage{epstopdf}
\usepackage{lineno}
\usepackage{cancel}

\usepackage{natbib}
\bibpunct{(}{)}{;}{a}{}{,}

\begin{document} 

\title{MHD study of planetary magnetospheric response during extreme solar wind conditions: Earth and exoplanet magnetospheres applications}

\titlerunning{Planetary magnetospheric response during extreme solar wind conditions}
\authorrunning{Varela et al.}

   \author{J. Varela\inst{1},
          A. S. Brun\inst{2}, 
          A. Strugarek\inst{2},  
          V. R\'eville\inst{3},   
          P. Zarka\inst{4},
               and  
          F. Pantellini\inst{5}
          }

   \institute{Universidad Carlos III de Madrid, Leganes, 28911 \\
              \email{\href{mailto:jvrodrig@fis.uc3m.es (telf: 0034645770344)}{jvrodrig@fis.uc3m.es}}
          \and  
             Laboratoire AIM, CEA/DRF – CNRS – Univ. Paris Diderot – IRFU/DAp, Paris-Saclay, 91191 Gif-sur-Yvette Cedex, France
          \and
            IRAP, Universit\'e Toulouse III—Paul Sabatier, CNRS, CNES, Toulouse, France
          \and
             LESIA \& USN, Observatoire de Paris, CNRS, PSL/SU/UPMC/UPD/UO, Place J. Janssen, 92195 Meudon, France
          \and
             LESIA, Observatoire de Paris, Universit\'e PSL, CNRS, Sorbonne Universit\'e, Universit\'e de Paris, 5 place Jules Janssen, 92195 Meudon, France  \\
             }

\date{version of \today}

 
  \abstract
  {
     \textit{Context:} The stellar wind and the interplanetary magnetic field modify the topology of planetary magnetospheres. Consequently, the hazardous effect of the direct exposition to the stellar wind, for example regarding the integrity of satellites orbiting the Earth or the habitability of exoplanets, depend upon the space weather conditions. \\
     \textit{Aims:} The aim of the study is to analyze the response of an Earth-like magnetosphere for various space weather conditions and interplanetary coronal mass ejections. The magnetopause stand off distance, open-close field line boundary and plasma flows towards the planet surface are calculated. \\
     \textit{Methods:} We use the MHD code PLUTO in spherical coordinates to perform a parametric study regarding the dynamic pressure and temperature of the stellar wind as well as the interplanetary magnetic field intensity and orientation. The range of the parameters analyzed extends from regular to extreme space weather conditions consistent with coronal mass ejections at the Earth orbit for the present and early periods of the Sun´s main sequence. In addition, implications of sub-Afvenic solar wind configurations for the Earth and exoplanet magnetospheres are analyzed..  \\
      \textit{Results:} The direct precipitation of the solar wind at the Earth day side in equatorial latitudes is extremely unlikely even during super coronal mass ejections. On the other hand, for early evolution phases along the Sun main sequence once the Sun rotation rate was at least $5$ times faster ($< 440$ Myr), the Earth surface was directly exposed to the solar wind during coronal mass ejections. Nowadays, satellites at High, Geosynchronous and Medium orbits are directly exposed to the solar wind during coronal mass ejections, because part of the orbit at the Earth day side is beyond the nose of the bow shock.
    }


\maketitle


\section{Introduction}

The space weather forecasting in the last decades has shown the important effect of the solar wind (SW) and interplanetary magnetic field (IMF) on the Earth magnetosphere, ionosphere, thermosphere and exosphere state \citep{Poppe,Gonzalez}. Physical phenomena as geomagnetic storms \citep{Gonzalez2} and substorms \citep{Baker3}, energization of the Van Allen radiation belts \citep{Shah}, ionospheric disturbances \citep{Cherniak}, aurora \citep{Zhang}, and geomagnetically induced currents at Earth's surface \citep{Pulkkinen} are triggered during particular space weather conditions. Extreme space weather conditions linked to coronary mass ejections (CME) lead to a strong perturbation of the Earth magnetosphere \citep{Cane,Richardson,Wang4,Lugaz,Wu}. The list of consequences is large: failure of spacecraft electronics due to radiation damage and charging \citep{Choi}, enhancement of the drag on low orbit satellites \citep{Nwankwo}, spacecraft signal scintillation due to a perturbed ionosphere \citep{Molera}, ground induced electric currents that can cause the collapse of electric power grids \citep{Cannon}, ionizing radiation that harms astronauts and passenger of the commercial aviation \citep{Bazilevskaya}, among others. Recently, the analysis of the space weather is generalized for the case of stars different than the Sun \citep{Strugarek,Garraffo}. Between other factors, the habitability of the exoplanets depends on the space weather conditions imposed by the hosting star and the shielding efficiency of the exoplanet magnetic field, avoiding the sterilizing effect of the stellar wind on the planet surface \citep{Gallet,Linsky,Airapetian}. In addition, the direct exposition of the exoplanet to the stellar wind leads to the depletion of the atmosphere, particularly volatile molecules as water by thermal and non-thermal escape \cite{Lundin,Moore,Jakosky}.

The CMEs are solar eruptions caused by magnetic reconnections in the star corona \citep{Low,Howard}, expelling a large amount of fast charged particles and a magnetic cloud that evolves into an interplanetary coronal mass ejection (ICME) \citep{Sheeley,Neugebauer,Cane2,Gosling}. If the ICME impacts the Earth, the measured SW dynamic pressure increases to $10 - 100 nPa$ and the IMF intensity to $100 - 300$ nT \citep{Gosling2,Huttunen,Manchester,Schwenn,Riley,Howard2,Mays,Kay,Savani,Salman,Kilpua,Hapgood}. The Disturbance Storm Time Index ($Dst$) indicates the magnetic activity derived from a network of near-equatorial geomagnetic observatories that measures the intensity of the globally symmetrical equatorial electrojet (the ring current), widely used to identify extreme SW / IMF space weather conditions \citep{Sugiura,Loewe,Siscoe,Borovsky}. A negative $Dst$ value means that Earth's magnetic field is weakened due to the IMF erosion, particularly during solar storms. The strongest event observed until the present days is the Carrington event that happened the year $1859$ \citep{Carrington}. An unusual large number of sunspots on the solar disk and a wide active region was registered from where an extremely fast ICME was launched toward the Earth. Several authors studied the Carrington event suggesting a shock traveling around $2000$ km/s \citep{Cliver} that generated the strongest geomagnetic storm with $Dst \approx -1700$ nT \citep{Tsurutani}, later revised to $Dst \approx -850$ nT by \citep{Siscoe}. The most recent strongest event, called Bastille day event ($14-16$ of July $2000$), leads to $Dst \approx -300$ nT for a SW velocity of $1000$ km/s and an IMF intensity of $\approx 45$ nT \citep{Rastatter}. On the other hand, typical ICMEs impacting the Earth shows an averaged plasma velocity of $350-500$ km/s and IMF intensities between $9-13$ nT leading to geomagnetic storms with $Dst < -50$ nT \citep{Cane2}.

The interaction of the SW with planetary magnetospheres can be studied using numerical models. Different computational frameworks were used, for example single fluid \citep{2008Icar..195....1K,2015JGRA..120.4763J,Strugarek2,Strugarek}, multifluid \citep{2008JGRA..113.9223K} and hydrid codes \citep{2010Icar..209...46W,Muller2011946,Muller2012666,2012JGRA..11710228R,Turc}. The simulations indicate a stronger compression of the bow shock as the SW dynamic pressure increases, as well as an enhancement or a weakening of the effective planet magnetic field according to the IMF orientation and intensity, leading to a modification of the magnetosphere topology \citep{Slavin,2000Icar..143..397K,2009Sci...324..606S}. Regarding the Earth magnetosphere, several MHD models were developed to analyze the interaction of the Earth magnetic field with the SW and IMF: GEDAS model \citep{Ogino2}, Tanaka model \citep{Tanaka}, Block-Adaptive Tree Solar-wind Roe-type Upwind Scheme (BATS-R-US) \citep{Powell}, Grand Unified Magnetosphere-Ionosphere Coupling Simulation, version 4 \citep{Janhunen}, Lyon-Fedder-Mobarry (LFM) model \citep{Lyon}, Space Weather Modelling Framework (SWMF) \citep{Toth}, Open General Geospace Circulation Model (OpenGGCM) \citep{Raeder}, Piecewise Parabolic Method with a Lagrangian Remap MHD (PPMLR-MHD) model \citep{Hu} and AMR-CESE-MHD model \citep{Wang5}. Thus, the effect of different SW and IMF configurations on the global structures of the Earth magnetosphere was already analyzed by several authors using MHD codes, particularly the Bow Shock \citep{Samsonov,Andreeova,Nemecek,Mejnertsen}, the Magnetosheath \citep{Ogino,Wang}, the magnetopause stand off distance \citep{Cairns,Cairns2,Wang2} and the magnetotail \citep{Laitinen,Wang3}. In addition, global MHD models were applied to analyze the interaction of ICMEs with the Earth magnetosphere \citep{Wu2,Wu3,Shen,Ngwira,Wu4,Scolini,Torok}. The simulations show large topological deformations caused by the combined effect of the SW dynamic pressure, IMF magnetic pressure and the reconnection between the IMF and the Earth magnetic field. Consequently, the magnetopause stand off distance significantly decreases \citep{Sibeck,Dusik,Liu,Nemecek2,Grygorov,Samsonov2}.

MHD codes were validated comparing the simulation results with ground based magnetometers and spacecraft measurements \citep{Watanabe}. For example, \citet{Raeder2} compared global Earth magnetosphere simulations with magnetometer and plasma data obtained from spacecrafts during the substorm event of $24/11/1996$. \citet{Wang6} calculated the plasma depletion layer and compared the results with WIND data. \citet{Den} developed a real-time Earth magnetosphere simulator using the data measured from the spacecraft ACE that was compared with geomagnetic field activities as well as real-time plasma temperature and density data at the geostationary orbit. \citet{Facsko} performed a one year global simulation of the Earth’s magnetosphere comparing the results with CLUSTER spacecraft measurements. In addition, predictions of BATS-R-US, the GUMICS, the LFM, and the OpenGGCM in \citet{Honkonen} were compared with the measurements of Cluster \citep{Escoubet}, WIND \citep{Acuna} and GEOTAIL \citep{Nishida} missions, as well as the Super Dual Auroral Radar Network (SuperDARN) \citep{Greenwald} cross polar cap potential (CPCP).

The aim of this study is to analyze the topology of the Earth magnetosphere and exoplanets with an Earth-like magnetosphere during coronal mass ejections. The study novelty lies in the extended use of parametric analysis to calculate the magnetosphere deformation trends regarding the SW and IMF properties. As new results, the study encompasses a forecast of the space weather conditions leading to the direct exposition of satellites to the SW at different orbits, as well as the direct precipitation of the SW towards the Earth / exoplanet surface. In addition, the shielding efficiency of the Earth magnetic field during the Sun evolution along the main sequence until the present day is analyzed, identifying the Sun evolution stage favorable to sustain life at the Earth surface considering both standard and extreme space weather conditions, assuming a fixed intensity of the Earth magnetic field. We also analyze the ICMEs that impacted the Earth from the year $1997$ to $2020$, particularly the response of the magnetosphere regarding the new ICME classification derived from our parametric study.

The present study is performed using the single fluid MHD code PLUTO in spherical 3D coordinates \citep{Mignone}. The analysis is based on an upgraded model previously applied in the study of the global structures of the Hermean magnetosphere \citep{Varela,Varela2,Varela3,Varela4,Varela5} and the radio emission from exoplanets \cite{Varela6}. In the present study, a set of simulations is performed with various dynamic pressure and temperature values of the SW as well as IMF intensities and orientations for the case of the Earth magnetosphere.

Single fluid MHD simulations cannot reproduce the kinetic process on planetary magnetospheres, leading to a deviation between simulation results and observations if the kinetic effects are large \citep{Chen,Aizawa}. Energy conversion processes \citep{Chaston}, ion range turbulence \citep{Chen2} between other examples are not correctly described by MHD simulations. This is also the case for the foreshock located upstream quasi-parallel bow shocks \citep{Omidi,Eastwood4}, linked to the formation of hot flow anomalies (HFAs) created by kinetic interactions between IMF discontinuities and the quasi-parallel bow shock \citep{Schwartz,Turner}, foreshock cavities showing low plasma density and magnetic strength as well as enhanced wave activity \citep{Katircioglu,Sibeck2} and foreshock bubbles generated during the interactions of counter-streaming suprathermal ions with IMF discontinuities \citep{Omidi2,Turner2}. The foreshock causes magnetosphere disturbances not reproduced by single fluid MHD models, thus kinetic \citep{Ilie,Chen3}, hybrid \citep{Lu,Lin} or multi-fluid \citep{Ma,Manuzzo} models are required for an improved concurrence of simulation results and observational data. Consequently, deviations could exist between present study simulation results and observational data for the case of extreme space weather configurations.

This paper is structured as follows. There is a description of the simulation model, boundary and initial conditions in section \ref{Model}. The distortion of the Earth magnetic field topology driven by the solar wind and interplanetary magnetic field is analyzed in section \ref{Topology}. The effect of the space weather conditions on the satellite integrity due to the direct exposition to the SW and the Earth habitability along the Sun main sequence are discussed in section \ref{Applications}. Finally, section \ref{Conclusions} shows the summary of the study main conclusions discussed in the context of other authors results.

\section{Numerical model}
\label{Model}

The simulations are performed using the ideal MHD version of the open source code PLUTO in spherical coordinates. The model solves the time evolution of a single fluid polytropic plasma in the non resistive and inviscid limit \citep{Mignone}. The equations solved in conservative form are:
\begin{equation}
\label{Density}
\frac{\partial \rho}{\partial t} + \mathbf{\nabla} \cdot \left( \rho \mathbf{v} \right) = 0
\end{equation}
\begin{equation} 
\label{Momentum}
\frac{\partial \mathbf{m}}{\partial t} + \mathbf{\nabla} \cdot \left[ \mathbf{m}\mathbf{v} - \frac{\mathbf{BB}}{\mu_{0}} + I \left( p + \frac{\mathbf{B}^{2}}{2\mu_{0}} \right)  \right]^{T} = 0 
\end{equation} 
\begin{equation}
\label{B field}
\frac{\partial \mathbf{B}}{\partial t} + \mathbf{\nabla} \times \left(\mathbf{E} \right) = 0
\end{equation} 
\begin{equation}
\label{Energy}
\frac{\partial E_{t}}{\partial t} + \mathbf{\nabla} \cdot \left[ \left( \frac{\rho \mathbf{v}^{2}}{2} + \rho e + p \right) \mathbf{v} + \frac{\mathbf{E} \times \mathbf{B}}{\mu_{0}}  \right] = 0
\end{equation}
$\rho$ is the mass density, $\mathbf{m} = \rho \mathbf{v}$ the momentum density, $\mathbf{v}$ the velocity, $p$ the gas thermal pressure, $\mathbf{B}$ the magnetic field, $E_{t} = \rho e + m^2/2 \rho + B^2/2\mu_{0}$ the total energy density,  $\mathbf{E} = -(\mathbf{v} \times \mathbf{B}) $ the electric field and $e$ the internal energy. The closure is provided by the equation of state $\rho e = p / (\gamma - 1)$ (ideal gas).

The conservative form of the equations are integrated using a Harten, Lax, Van Leer approximate Riemann solver (hll) associated with a diffusive limiter (minmod). The initial magnetic fields are divergenceless, condition maintain towards the simulation by a mixed hyperbolic/parabolic divergence cleaning technique \citep{Dedner}.

The grid is made of $128$ radial points, $48$ in the polar angle $\theta$ and $96$ in the azimuthal angle $\phi$. The grid is equidistant in the radial direction and the cell volume increases beyond the inner domain of the simulation. The simulation domain is defined as two concentric shells around the planet with $R_{in} = 2R_{E}$ the inner boundary ($R_{in} = 3R_{E}$ if the SW dynamic pressure is smaller than $1$ nPa) and $R_{out} = 30R_{E}$ the outer boundary, with $R_{E}$ the Earth radius. The simulation characteristic length is $L =6.4 \cdot 10^{6}$ m (the Earth radius), $V = 10^{5}$ m/s the simulation characteristic velocity (order of magnitude of the solar wind velocity), the numerical magnetic diffusivity $\eta \approx 5 \cdot 10^{8}$ m$^{2}/s$ and the numerical kinematic diffusivity $\nu \approx 10^{9}$ m$^{2}$/s, thus the effective numerical magnetic Reynolds number due to the grid resolution is $R_{m}= V L/\eta \approx 1280$ and the kinetic Reynolds number $R_{e}=V L/\nu \approx 640$ (magnetic Prandtl number $P_{m} = R_{m}/R_{e}=2$). No explicit value of the dissipation is included in the model, hence the numerical magnetic diffusivity regulates the typical reconnection in the slow (Sweet–Parker model) regime. There is a detail discussion of the numerical magnetic and kinetic diffusivity of the model in \citep{Varela6}.

An upper ionosphere model is introduced between $R_{in}$ and $R = 2.5R_{E}$ where special conditions apply ($R_{in} = 3.0$ and $3.5R_{E}$ if the SW dynamic pressure is smaller than $1$ nPa). The upper ionosphere model is described in the Appendix A, based on the electric field generated by the field aligned currents providing the plasma velocity at the upper ionosphere. The outer boundary is divided in the upstream part where the stellar wind parameters are fixed and the downstream part where the null derivative condition $\frac{\partial}{\partial r} = 0$ for all fields is assumed. Regarding the initial conditions of the simulations, the IMF is cut off at $R_{c} = 8R_{E}$. In addition, a paraboloid with the vertex at the day side of the planet is defined as $x < A - (y^2 + z^{2} / B)$, with $(x,y,z)$ the Cartesian coordinates, $A = R_{c}$ and $B = R_{c}*\sqrt{R_{c}}$ where the velocity is null and the density profile is adjusted to keep the Alfv\'{e}n velocity constant $\mathrm{v}_{A} = B / \sqrt{\mu_{0}\rho} = 8 \cdot 10^{3}$ km/s with $\rho = nm_{p}$ the mass density, $n$ the particle number and $m_{p}$ the proton mass. It should be noted that, $\mathrm{v}_{A} \approx 10^{4}$ km/s corresponds to a Alfv\'{e}n velocity $2-3$ smaller with respect to the Alfv\'{e}n velocity at $R = 2.5R_{E}$ \citep{Run}, required to keep a time step large enough for the simulation to remain tractable.

The Earth magnetic field is implemented as a dipole rotated $90^{0}$ in the YZ plane with respect to the grid poles. In this way, the magnetic field do not correspond to the grid poles avoiding numerical issues, thus no special treat is included for the singularity at the magnetic poles. The effect of the tilt of the Earth rotation axis with respect to the Ecliptic plane ($23^{o}$) is emulated modifying the orientation of the IMF and stellar wind velocity vectors (no dipole tilt is included for simplicity, thus the geographical and magnetic poles are the same). The simulation frame is such that the z-axis is given by the planetary magnetic axis pointing to the magnetic North pole and the star-planet line is located in the XZ plane with $x_{star} > 0$ (Solar Magnetospheric coordinates). The y-axis completes the right-handed system. 

The model assumes a fully ionized proton electron plasma. The sound speed is defined as $\mathrm{c} = \sqrt {\gamma p/\rho} $ (with $p$ the total electron + proton pressure and $\gamma=5/3$ the adiabatic index), the sonic Mach number as $M_{s} = \mathrm{v}/\mathrm{c}$ and the Alfv\'{e}nic Mach number as $M_{a} = \mathrm{v}/\mathrm{v}_{A}$, with $\mathrm{v}$ the plasma velocity. It should be noted that, the present model does not resolve the plasma depletion layer as a decoupled global structure from the magnetosheath due to the lack of model resolution. Nevertheless, the model is able to reproduce the global magnetosphere structures as the magnetosheath and magnetopause, as it was demonstrated for the case of the Hermean magnetosphere \citep{Varela,Varela2,Varela3}. In addition, the reconnection between interplanetary and Earth magnetic field is instantaneous (no magnetic pile-up on the planet dayside) and stronger (enhanced erosion of the planet magnetic field) because the magnetic diffusion of the model is larger with respect to the real plasma, although the effect of the reconnection region on the depletion of the magnetosheath and the injection of plasma into the inner magnetosphere is correctly reproduced in a first approximation. Also, the Earth rotation and orbital motion is not included in the model yet and let for future work.

Our subset of ICME simulations aims at computing the Earth magnetosphere topology for the largest forcing caused by the space weather conditions, reason why the simulation input is selected once the local maxima of dynamic pressure, IMF intensity and Southward IMF component is reached, see Appendix D for details. Nevertheless, there is a relaxation time required by the Earth magnetosphere to evolve between different configurations if the space weather conditions change. The magnetosphere relaxation time due to variations of the IMF orientation and intensity is linked to the reconnection rate with the Earth magnetic field, analyzed in detail by \citep{Borovsky2,Burch}. A response time of around $6$ min was measured by the Magnetospheric Multiscale Science (MMS) satellite \citep{Fuselier} for the reconnection region during a Northward inversion of the IMF \citep{Trattner}. In addition, the study by \citet{Trattner} indicates that slow changes in the IMF lead to a fast response time with respect to the reconnection location, although rapid changes lead to a delay of several minutes in the reconnection location response. Also, simulations by \citet{Zeeuw} calculated an answer time of around $10$ min for the subauroral ionospheric electric field after a Northward IMF inversion. The relaxation time and magnetosphere dynamics due to variations of the SW dynamic pressure and temperature were analyzed by \citet{Eastwood3,Zhang2,Nishimura,Shi}, showing a large variety of transient events that can last from seconds to a hundred of minutes. Consequently, several response times exist linked to different magnetospheric processes, although in the present study the main response time is the relaxation time required by the dayside magnetopause to reach a new equilibrium position, linked to the time required by the Alfv\'en wave to travel a distance of the order of the magnetopause standoff distance (Alfv\'en crossing time). The evolution of the space weather conditions could be very fast during the impact of the ICME, leading to inversions of the IMF components as well as local peaks of the SW dynamic pressure and temperature in a few minutes. Thus, the relaxation time could be exceeded and the Earth magnetosphere topology shows a memory regarding previous configurations. Consequently, the simulations performed could overestimate the forcing of the SW and IMF because the effect imprinted in the Earth magnetosphere by previous space weather conditions are not considered.

The magnetosphere response to the SW and IMF show several interlinked phases that must be distinguished. First, the response of the day side magnetopause and magnetosheath affecting the magnetosphere stand off distance, plasma flows toward the inner magnetosphere or the location of the reconnection regions, between other consequences. Next, the response of the magnetotail, followed by the ionospheric response and subsequently the ring current response. It should be noted that the analysis is mainly dedicated to the day side response of the magnetosphere. The analysis of the magnetotail is not performed in detail, although some implications regarding the magnetic field at the night side are discussed. On the other hand, the response of the ionosphere and ring current are out of the scope of the study.

The IMF and SW parameters are fixed, that is to say, the simulation is assumed complete once the steady state is reached. Thus, dynamic events caused by the evolving space weather conditions are not included in the study. The simulations reach the steady state after $\tau = L / V = 15$ code time, equivalent to $t \approx 16$ min of Physical time, although the magnetosphere topology in the Earth day side is steady after $t \approx 11$ min. Consequently, the code can reproduce accurately the magnetosphere response if the variation of the space weather conditions are roughly steady for time periods of $t = 10-15$ min.

The study includes the analysis of the space weather during normal, CME and super-CME conditions. Table \ref{1} shows the parameter range for each space weather condition:

\begin{table}[h]
\centering
\begin{tabular}{c | c c c c}
Case & $n$ & $|v|$ & $T$ & $|B|_{IMF}$\\
 & (cm)$^{-3}$ & (km/s) & ($10^{3}$ K) & (nT) \\ \hline
Normal & $\le 10$ & $< 500$ & $< 60$ & $\le 10$  \\
CME & $[10 , 120]$ & $[500 , 1000]$ & $[60 , 200]$ & $[10 , 100]$ \\
S-CME & $> 120$ & $> 1000$ & $> 100$ & $> 100$ \\
\end{tabular}
\caption{Space weather classification with respect to the SW density, velocity and temperature as well as the IMF intensity.}
\label{1}
\end{table}

The range of SW and IMF parameters explored in this study exceeds the present space weather condition for the Earth. The most extreme configurations show the space weather conditions that could exist during an early period of the Sun main sequence or for the case of an exoplanet magnetosphere. Appendix F includes the list of SW and IMF parameters used in the different analysis performed in section \ref{Topology}. 

In addition, the effect of six different IMF orientations are considered in the study: Earth-Sun and Sun-Earth (also called radial IMF configurations), Southward, Northward, Ecliptic clockwise and Ecliptic counter clockwise. Earth-Sun and Sun-Earth configurations indicate an IMF parallel to the SW velocity vector. Southward and Northward IMF orientations show an IMF perpendicular to the SW velocity vector at the XZ plane. Consequently, because the tilt of the Earth rotation axis with respect to the ecliptic plane is included in the model, the simulations show a North-South asymmetry of the magnetosphere.

\section{Effect of the SW and IMF on the Earth / exoplanet magnetosphere topology}
\label{Topology}

Figure \ref{1} shows a 3D view of the system for a Northward IMF orientation. There is an accumulation of plasma at the planet day side because the SW is slowed down and diverged due to the interaction with the planet magnetic field, thus the Bow Shock (BS) in the simulations is identified as the region showing a sudden increase of the plasma density ($5$ times larger with respect to the SW density). The SW dynamic pressure bends the planet magnetic field lines (red lines), compressed on the planet day side and stretched at the nigh side forming the magnetotail. In addition, the planet magnetic field lines reconnect with the IMF leading to a local erosion/enhancement of the magnetosphere. The yellow arrows indicate the IMF orientation and the dashed white line the outer limit of the simulation domain (the star is not included in the model). It should be noted that the magnetotail can extend more than $100R_{E}$ although the computation domain is limited to $30R_{E}$, thus the model only reproduces partially this magnetosphere structure if the SW dynamic pressure is $\ge 50$ nPa and the IMF intensity is $\le 10$ nT. A detail discussion is done in the Appendix C.

\begin{figure}[h]
\centering
\resizebox{\hsize}{!}{\includegraphics[width=\columnwidth]{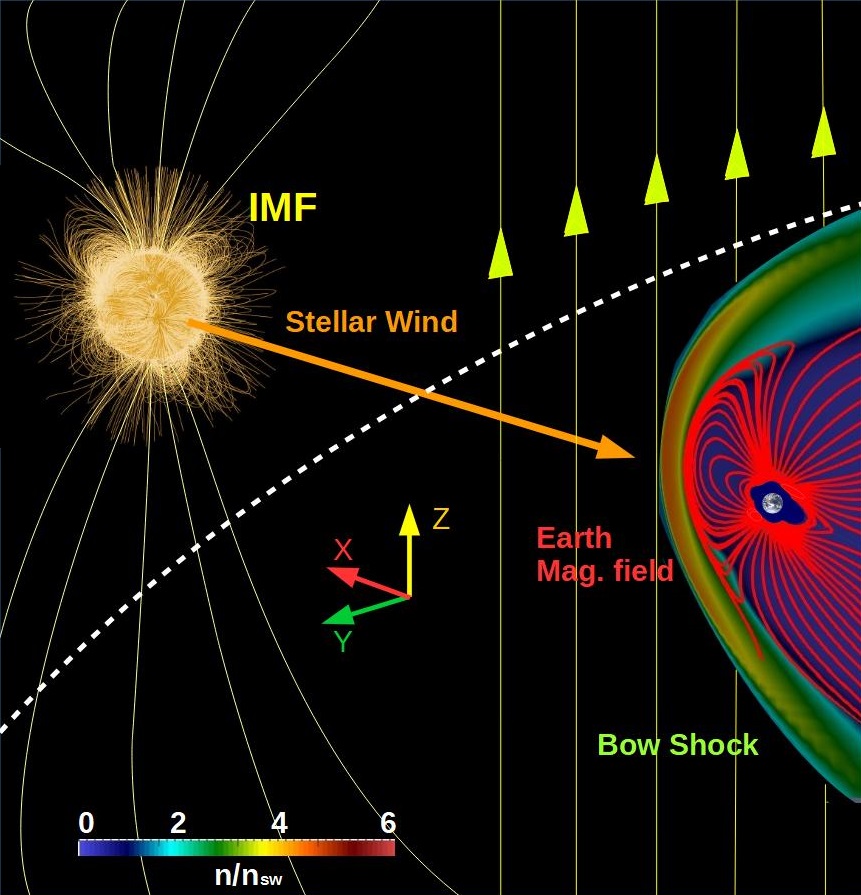}}
\caption{3D view of a typical simulation setup. Density distribution (color scale), Earth magnetic field lines (red lines) and IMF (yellow lines). The yellow arrows indicate the orientation of the IMF (Northward orientation). The dashed white line shows the beginning of the simulation domain (note that the star is not included in the model).}
\label{1}
\end{figure}

Figure \ref{2} illustrates the effect of the IMF showing the planet magnetic field (red lines), SW stream lines (green lines), reconnection region ($|B| = 10$ nT isocontour of the magnetic field, pink lines), the nose of the BS ($v_{r} = 0$ isocontour, white lines) and the regions where the magnetosheath plasma is injected into the magnetosphere (bold cyan arrows) in the XY plane. We should clarify that the definition of the magnetosphere reconnection regions is given by the antiparallel reconnection model, that is to say, the regions with antiparallel magnetic fields. The simulations are performed for different IMF orientations, IMF intensities and dynamic pressure values. In the following, the discussion of the simulation results refers only to the Earth magnetosphere for simplicity, even though some of the configurations analyzed do not correspond to the present space weather conditions. Such special configurations are highlighted to avoid misunderstanding.

\begin{figure}[h]
\centering
\includegraphics[width=8cm]{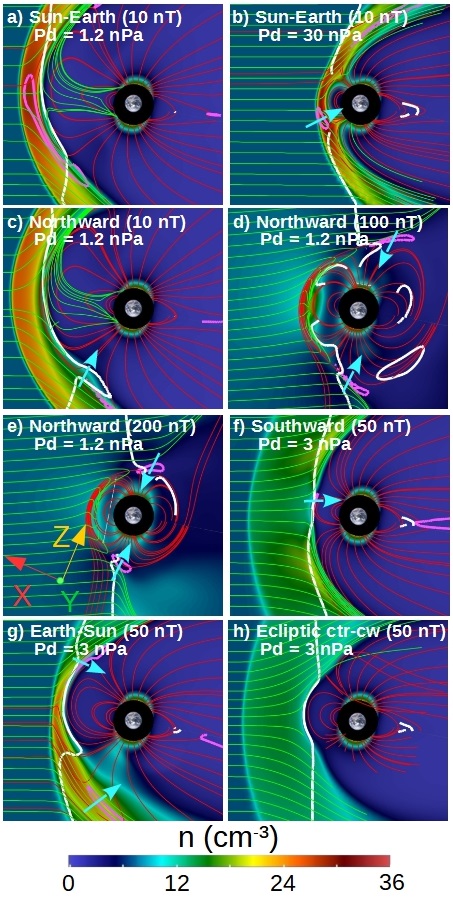}
\caption{Polar cut (XY plane) of the plasma density in simulations with (a) Sun-Earth IMF orientation $|B_{IMF}| = 10$ nT $P_{d} = 1.2$ nPa, (b) Sun-Earth IMF orientation $|B_{IMF}| = 10$ nT $P_{d} = 30$ nPa, (c) Northward IMF orientation $|B_{IMF}| = 10$ nT $P_{d} = 1.2$ nPa, (d) Northward IMF orientation $|B_{IMF}| = 100$ nT $P_{d} = 1.2$ nPa, (e) Northward IMF orientation $|B_{IMF}| = 200$ nT $P_{d} = 1.2$ nPa, (f) Southward IMF orientation $|B_{IMF}| = 50$ nT $P_{d} = 3$ nPa, (g) Earth-Sun IMF orientation $|B_{IMF}| = 50$ nT $P_{d} = 3$ nPa and (h) Ecliptic ctr-cw IMF orientation $|B_{IMF}| = 50$ nT $P_{d} = 3$ nPa. Earth magnetic field (red lines), SW stream functions (green lines), $|B| = 10$ nT isocontour of the magnetic field (pink lines) and $v_{r} = 0$ isocontours (white lines). The bold white arrows shows the regions where the plasma is injected into the inner magnetosphere.}
\label{2}
\end{figure}

The simulations show a stronger compression of the magnetosphere as the dynamic pressure increases leading to a smaller magnetopause stand off distance, see panels a and b. The simulations also shows a large deformation of the Earth magnetosphere if $|B_{IMF}|$ increases. For example, if $|B_{IMF}|$ increases from $10$ to $200$ nT for a Northward IMF orientation, see panels c to e, the reconnection region between the IMF and the Earth magnetic field is located closer to the poles, enhancing the plasma flows towards the Earth poles. Consequently, the IMF modifies the plasma injection into the inner magnetosphere, and therefore the plasma flows towards the Earth surface along the magnetic field lines (bold white arrows). In addition, the magnetosphere is compressed in the magnetic axis direction and the magnetopause stand off distance decreases. On the other hand, Southward IMF orientations lead to a magnetic reconnection in the equatorial region that erodes the Earth magnetic field, causing a decrease of the magnetopause stand off distance and the injection of SW in the inner magnetosphere at a lower latitude, see panel f. Furthermore, the Earth-Sun (Sun-Earth) IMF orientation causes a Northward (Southward) displacement at the day side (DS) and a Southward (Northward) displacement at the night side (NS), see panels a and g. Finally, a IMF orientation in the Ecliptic plane causes an East/West tilt of the Earth magnetosphere. It should be noted that the simulations with a SW density of $12$ cm$^{-3}$ and $|B|_{IMF} \le 60$ nT lead to $M_{a} < 1$ ($\mathrm{v}_{A} = 378$ km/s if $|B_{IMF}| = 60$ nT) thus the BS is not formed, consistent with the observations by \citet{Lavraud,Chane,Lugaz2}. This is the case of the simulations shown in the panels d and e.

The deformations induced by the SW / IMF in the Earth magnetosphere during extreme space weather conditions are very large. Figure \ref{3} show some examples of extreme weather conditions regarding the IMF intensity, 3d views of the Earth magnetosphere if $|B|_{IMF} = 250$ nT and $P_{d} = 1.2$ nPa for different IMF orientations. The panel (a) indicates a simulation with Sun-Earth IMF, panel (b) Southward IMF, panel (c) Northward IMF and panel (d) ecliptic ctr-clockwise IMF.

\begin{figure}[h]
\centering
\resizebox{\hsize}{!}{\includegraphics[width=\columnwidth]{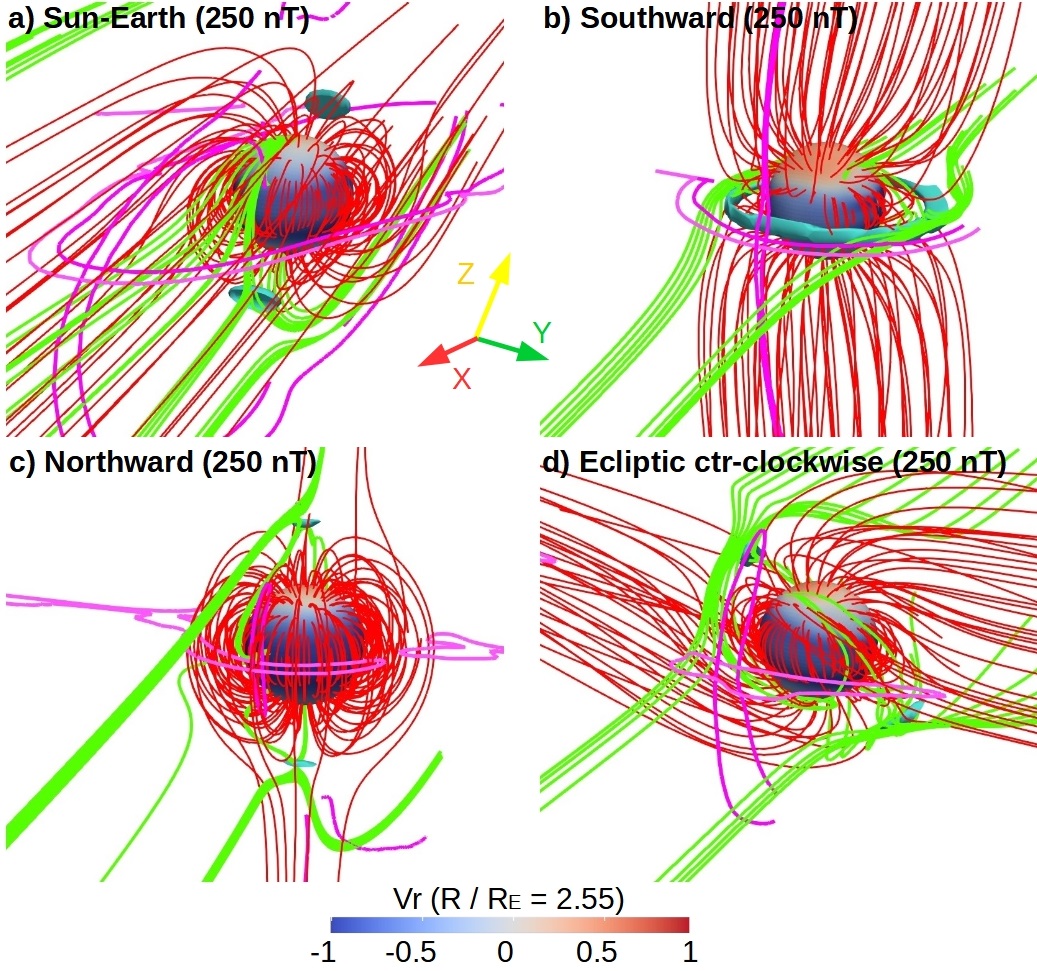}}
\caption{3D view of the Earth magnetosphere topology if $|B|_{IMF} = 250$ nT for (a) a Sun-Earth, (b) Southward, (c) Northward and (d) ecliptic ctr-clockwise IMF orientations. Earth magnetic field (red lines), SW stream functions ( green lines) and isocontours of the plasma density for $6 - 9$ cm$^{-3}$ indicating the location of the BS (pink lines). The blue isocontours indicate the reconnection regions ($|B| = 60$ nT).}
\label{3}
\end{figure}

The simulations show that the reconnection regions (blue isocontour of the magnetic field) and the BS (pink lines of the density isocontour cut with the XZ and XY planes) are located close to Earth surface (slightly above $R/R_{E} = 3$), pointing out the decrease of the magnetopause stand off distance with respect to the simulation with a weaker $|B_{IMF}|$. The $|B_{IMF}|$ during the impact of an ICME with the Earth is generally limited to $|B|_{IMF} < 100$ nT, thus space weather conditions with $|B_{IMF}| = 250$ nT falls in the category of super-ICMEs. The simulations indicate how the plasma is injected inside the inner magnetosphere through the reconnection regions, flowing among the Earth magnetic field lines from the magnetosheath towards the planet surface (green lines connected with inflow regions at $R/R_{E = 2.55}$, blue colors). If the IMF is Sun-Earth oriented, the Southward bending of the magnetosphere at the Earth day side enhances the plasma flows towards the North pole. The Southward IMF erodes the Earth magnetic field at the Ecliptic plane thus the plasma flows towards the Equator increase. On the other hand, the Northward IMF erodes the Earth magnetic field near the magnetic axis promoting the plasma flows towards the Poles. Furthermore, the Ecliptic IMF orientation induces a West/East tilt in the magnetosphere tilt and the plasma flows towards higher longitudes.

The simulations, after reaching the steady state, show the formation of a low density and high temperature plasma belt above the upper ionosphere. The plasma belt, trapped inside the closed magnetic field lines of the Earth, is generated from two main sources: the solar wind injected into the inner magnetosphere toward the reconnection regions and a plasma outward flux from the upper ionosphere to the simulation domain, see figure \ref{4}. The plasma belt in the simulations shares some features with the Van Allen radiation belt \citep{Allen,LiW} and the Earth's ring current \citep{Daglis,Ganushkina}, although it lacks the complexity of the real magnetosphere structures that cannot be reproduced by a single fluid MHD model \citep{Hudson,Kress,Jordanova}. In addition, the plasma belt narrows as the magnetopause stand off distance decreases, and is not observed in simulations that reproduce extreme space weather conditions (the plasma belt is located below $R/R_{E} = 2.5$). Likewise, other magnetosphere region as the plasmasphere cannot be correctly reproduced \citep{Singh}. Consequently, the analysis of the plasma belt, ring current and plasmasphere are out of the scope of the present study. It should be noted that these model limitations can lead to deviations between the simulation results and the observational data during extreme space weather conditions. 

\begin{figure}[h]
\centering
\resizebox{\hsize}{!}{\includegraphics[width=\columnwidth]{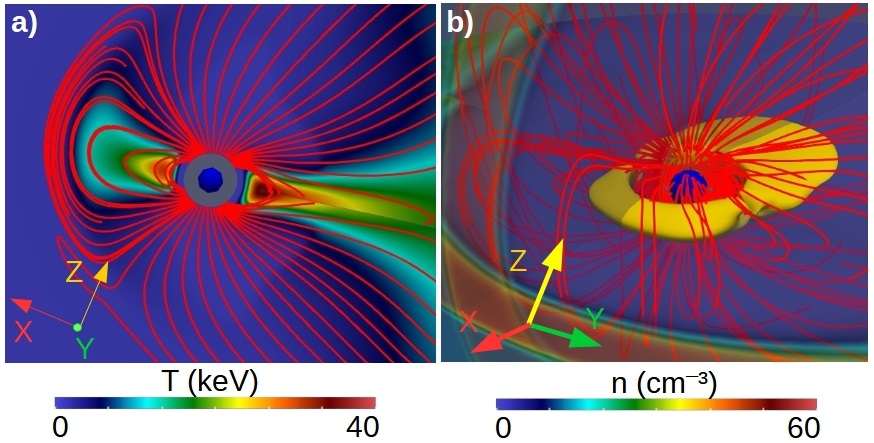}}
\caption{(a) Polar cut (XY plane) of the plasma temperature and (b) 3D view of the Earth magnetosphere adding the plasma temperature isocountour $T=26$ keV (orange surface, temperature local maxima at $R = 3 R_{E}$ planet day side) and a polar/equatorial (XY / XZ plane) cut of the plasma density for a simulation with no IMF and $P_{d} = 1.2$ nPa. The red lines indicate the Earth magnetic field lines.}
\label{4}
\end{figure}

Summarizing, the correct characterization of the Earth magnetosphere topology with respect to the IMF intensity and orientation requires a detailed parametric study for regular and extreme space weather conditions. Such analysis is performed in the following sections, dedicated to calculate the magnetopause stand off distance, the location of the reconnection regions and the open-close field line boundary for different IMF intensities and orientations.

\subsection{Parametric study of the magnetopause stand off distance}

The magnetopause stand off distance $R_{sd}$ could be calculated as the location where the dynamic pressure of the SW ($P_{d} = m_{p} n_{sw}  v_{sw}^{2}/2$), the thermal pressure of the SW ($P_{th,sw} = m_{p} n_{sw}  v_{th,sw}^{2}/2 = m_{p} n_{sw} c_{sw}^{2}/\gamma$) and the magnetic pressure of the IMF ($P_{mag,sw} = B_{sw}^{2}/(2 \mu_{0})$ are balanced by the magnetic pressure of the Earth magnetosphere of a dipolar magnetic field ($P_{mag,E} = \alpha \mu_{0} M_{E}^{2} / 8 \pi^2 r^{6} $) and the thermal pressure of the magnetosphere ($P_{th,MSP} = m_{p} n_{MSP}  v_{th,MSP}^{2}/2$), resulting into the expression:
\begin{equation}
\label{eqn:pressure}
P_{d} + P_{mag,sw} + P_{th,sw} = P_{mag,E} + P_{th,MSP}
\end{equation}
\begin{equation}
\label{eqn:balance}
\frac{R_{sd}}{R_{E}} = \left[ \frac{\alpha \mu_{0} M_{E}^{2}}{4 \pi^2 \left( m_{p} n_{sw}  v_{sw}^{2} + \frac{B_{sw}^{2}}{\mu_{0}} + \frac{2 m_{p} n_{sw}  c_{sw}^{2}}{\gamma} - m_{p} n_{BS} v_{th,MSP}^{2} \right)} \right]^{(1/6)}
\end{equation}
with $M_{E}$ the Earth dipole magnetic field moment, $r = R_{sd} / R_{E}$ and $\alpha$ the dipole compression coefficient ($\alpha \approx 2$ \citep{Gombosi}). This expression is an approximation and it does not consider the effect of the reconnection between the Earth magnetic field with the IMF, that is to say, the approximation assumes a compressed dipolar magnetic field ignoring the orientation of the IMF. Consequently, the theoretical stand off distance is only valid if $|B_{IMF}|$ is small, thus $R_{sd}/R_{E}$ should be calculated using simulations for extreme space weather conditions. In the following, the location of the magnetopause is defined as the last close magnetic field line at the Earth day side at $0^{o}$ longitude in the ecliptic plane. Figure \ref{5} shows the pressure balance in simulations without IMF and low $P_{d}$, large $|B_{IMF}|$ and low $P_{d}$ as well as large $|B_{IMF}|$ and large $P_{d}$.

\begin{figure*}[h]
\centering
\resizebox{\hsize}{!}{\includegraphics[width=\columnwidth]{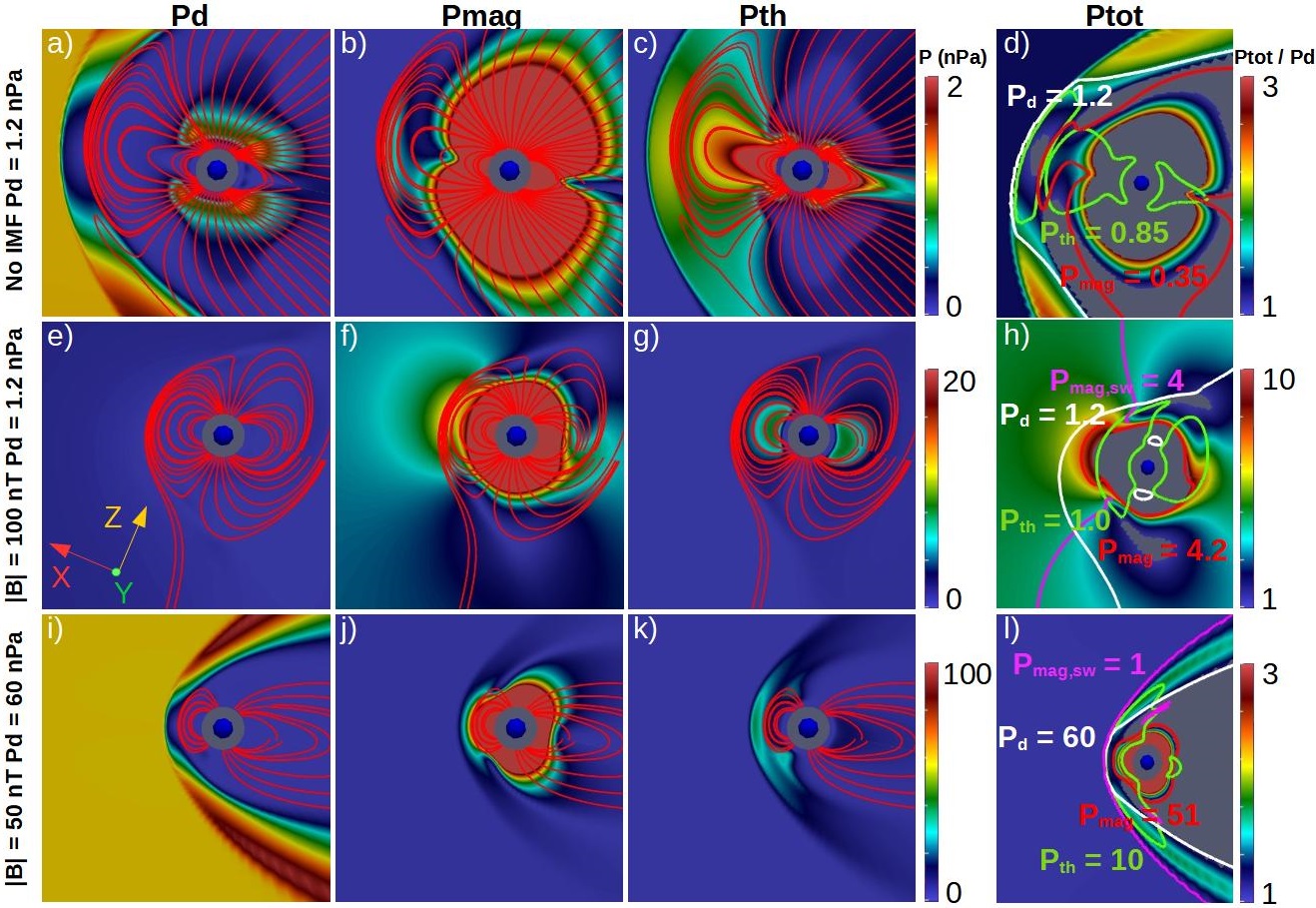}}
\caption{Polar cut (XY plane) of the pressure balance. Simulation with no IMF and $P_{d} = 1.2$ nPa, isocontour of (a) $P_{d}$, (b) $P_{mag}$ and (c) $P_{th}$. Simulation with Northward $|B_{IMF}| = 100$ nT and $P_{d} = 1.2$ nPa, isocontour of (e) $P_{d}$, (f) $P_{mag}$ and (g) $P_{th}$. Simulation with Northward $|B_{IMF}| = 50$ nT and $P_{d} = 60$ nPa, isocontour of (i) $P_{d}$, (j) $P_{mag}$ and (k) $P_{th}$. Panels (d), (h) and (l) show the total pressure ($P_{tot} = P_{d} + P_{mag} + P_{mag,sw} + P_{th}$) normalized to the SW dynamic pressure (isocontour) as well as the isolines of $P_{d}$ (white line), $P_{th}$ (green line), $P_{mag}$ (red line) and $P_{mag,sw}$ (pink line), including the respective isoline values (colored characters).}
\label{5}
\end{figure*}

The simulation without IMF and $P_{d} = 1.2$ nPa shows a balance between the dynamic pressure of the SW and the combined effect of the magnetosphere magnetic and thermal pressure, see panel a to d of fig \ref{5}. The effect of the magnetosphere thermal pressure is important on the pressure balance for space weather conditions with low $|B_{IMF}|$ and $P_{d}$, leading to $P_{th,MSP} / P_{mag,E} \approx 1.0$. It should be noted that fig \ref{5}, panels (c) and (d), show two local maxima of $P_{th}$ inside the BS and nearby the upper ionosphere. The $P_{th}$ local maxima nearby the upper ionosphere is linked to the plasma belt (see fig \ref{4}), which role on the pressure balance is negligible because the magnetic pressure generated by the Earth magnetic field in this plasma region is dominant, at least one order of magnitude higher. For a Northward IMF with $|B_{IMF}| = 100$ nT and $P_{d} = 1.2$ nPa, the leading terms in the pressure balance are the magnetic pressure of the IMF ($P_{d}$ is $3.5$ times smaller) and the magnetosphere magnetic pressure (the magnetosphere thermal pressure is $4$ times smaller), see panels e to h. Consequently, the IMF orientation is particularly important for space weather conditions with large IMF intensity although low SW dynamic pressure. On the other hand, the simulation for a Northward IMF with $|B_{IMF}| = 50$ nT and $P_{d} = 60$ nPa indicates a balance between the magnetic pressure of the magnetosphere (the magnetosphere thermal pressure is $4-5$ times smaller) and the combined effect of the SW dynamic pressure and the IMF magnetic pressure, see panels i to l. In other words, the leading terms of the pressure balance during extreme space weather conditions are the dynamic pressure of the SW, IMF magnetic pressure and the magnetosphere magnetic pressure.

We now turn to study the effect of the IMF intensity and orientation on the magnetopause stand off distance. For this purpose, we will fix the SW parameters to $T_{sw} = 1.8 \cdot 10^{5}$ K and $P_{d} = 1.2$ nPa. First, we must clarify that the configurations analyzed are idealizations, that is to say, an IMF purely oriented towards one direction is rarely observed particularly if the IMF intensity is large. This subtlety specially applies to the radial IMF configurations, because small deviations on the ecliptic component breaks the East-West symmetry on the model leading to a substantial variation of the Earth magnetosphere topology. Nevertheless, all the possible configurations are analyzed for the completeness of the study, independently of the rarity of the space weather condition. Figure \ref{6} shows the location of the magnetopause in the ecliptic plane for different IMF orientations and intensities.

\begin{figure}[h]
\centering
\resizebox{\hsize}{!}{\includegraphics[width=\columnwidth]{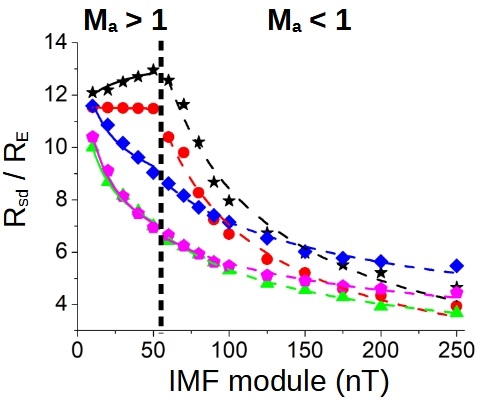}}
\caption{Magnetopause stand off distance with respect to $|B|_{IMF}$ if the IMF is oriented in the Sun-Earth direction (black star), Earth-Sun (red circle), Northward (blue diamond), Southward (green triangle) and Ecliptic ctr-clockwise direction (pink hexagon). The solid (dashed) lines indicate, for each IMF orientation, the data fit to the expression $R_{sd}/R_{E} = A |B|_{IMF}^{\alpha}$ of the simulations with $M_{a} > 1$ ($M_{a} < 1$).}
\label{6}
\end{figure}   

Two different trends are observed in fig \ref{6} for $R_{sd} / R_{E}$ regarding the $M_{a}$ value of the simulation. If $M_{a} < 1$, simulations with $|B_{IMF}| \le 60$ nT, the pressure balance is dominated by the magnetic pressure of the IMF and the Earth magnetic field because the BS is not formed, thus the thermal pressure of the plasma inside the BS does not participate to the balance, see fig \ref{2} panels d and e as well as fig \ref{5} panels d to f. On the other hand, if $M_{a} > 1$, the thermal pressure of the plasma inside the BS participates in the balance, particularly in the simulations with small $|B_{IMF}|$ values, see fig \ref{2} panels a and c as well as fig \ref{5} panels a to d ($P_{th,MSP} / P_{mag,E} \approx 0.4 - 1.0$). The general trend in the simulations with $M_{a} < 1$ indicates a decrease of $R_{sd}/R_{E}$ as the IMF intensity increases for all the IMF orientations. On the contrary, $M_{a} > 1$ simulations for the Sun-Earth and Earth-Sun IMF orientations show an increase or a constant $R_{sd}/R_{E}$, respectively. This exception is explained by the Northward (Southward) bending of the magnetosphere at the planet day side if the IMF is Earth-Sun (Sun-Earth), see fig \ref{2} panel g, as well as the magnetosphere thermal pressure. The IMF orientation that leads to the lowest $R_{sd}/R_{E}$ as $|B_{IMF}|$ increases is the Southward orientation while the Northward IMF orientation leads to the largest $R_{sd}/R_{E}$. The data for each IMF orientation and $M_{a}$ trend is fitted to the expression $R_{sd}/R_{E} = A |B|_{IMF}^{\alpha}$, indicated by solid lines for the simulations with $M_{a} > 1$ and dashed line for the $M_{a} < 1$ simulations in figure \ref{6}. Table \ref{2} shows the fitting parameters of the regressions in the simulations with $M_{a} < 1$ and $M_{a} > 1$ for different IMF orientations.
  
\begin{table}[h]
\centering
\begin{tabular}{c c c}
IMF & No BS ($M_{a} < 1$) & BS ($M_{a} > 1$) \\
\end{tabular}
\begin{tabular}{c | c c | c c}
 & A & $\alpha$ & A & $\alpha$ \\ \hline
Sun-Earth & $220$ & $-0.71$ & $10.9$ & $0.043$ \\
 & $\pm 40$ & $\pm 0.04$ & $\pm 0.3$ & $\pm 0.008$ \\
Earth-Sun & $210$ & $-0.73$ & $11.608$ & $-0.0028$ \\
 & $\pm 30$ & $\pm 0.03$ & $\pm 0.014$ & $\pm 0.0004$ \\ 
Northward & $35.1$ & $-0.345$ & $16.5$ & $-0.146$ \\
 & $\pm 0.9$ & $\pm 0.005$ & $\pm 0.9$ & $\pm 0.017$ \\
Southward & $33.9$ & $-0.402$ & $16.3$ & $-0.209$ \\
 & $\pm 1.4$ & $\pm 0.009$ & $\pm 0.7$ & $\pm 0.014$ \\
Ecliptic & $22.2$ & $-0.300$ & $18.5$ & $-0.244$ \\
 & $\pm 1.3$ & $\pm 0.013$ & $\pm 0.9$ & $\pm 0.016$ \\ \hline
\end{tabular}
\caption{Fit parameters of the regression $R_{sd}/R_{E} = A |B|_{IMF}^{\alpha}$ for different IMF orientations (first column) in simulations with $M_{a} < 1$ (second and third columns) and $M_{a} > 1$ (fourth and fifth columns).The standard errors of the regression parameters are included.}
\label{2}
\end{table}

The fit exponent of the simulations with $M_{a} < 1$ for Northward, Southward and Ecliptic IMF orientations are close to the theoretical $\alpha =-0.33$ value from the equation (6) neglecting the effect of the SW thermal and dynamic pressure as well as the magnetosphere thermal pressure. The excursion from the theoretical value is consequence of the IMF orientation, that is to say, due to the deviation from the dipolar magnetic field assumption. The largest deviation is observed for the Southward IMF orientation, because the Southward IMF leads to the strongest erosion of the Earth magnetic field at the day side and the largest decrease of the magnetopause stand off distance. It should be noted that the exponents are negative because the magnetic pressure of the IMF opposes to that of the magnetic pressure of the Earth magnetic field. On the other hand, the fit exponents for the Earth-Sun and Sun-Earth IMF orientations are more than $2$ times larger regarding the theoretical value. The large deviation is explained by the formation of two Alfv\'{e}n wings at the Earth day and night side \citep{Chane,Chane2}. Fig \ref{7}, panel a, shows the Alfv\'{e}n wings formed in the simulation with Earth-Sun IMF $|B_{IMF}| = 250$ nT and $P_{d} = 1.2$ nPa. The Alfv\'{e}n wings show the characteristic bending of the Earth magnetic field near the planet surface, the low velocity plasma inside the wings and a high velocity plasma linked to the reconnection regions between the IMF (white lines) and the Earth magnetic field (red lines). The IMF and Earth magnetic field magnetic pressure, see fig \ref{7} panel b, illustrates the role of the reconnection regions in the pressure balance and explains the large deviation of the fit exponents from the theoretical value. It must be pointed out that the Alfv\'{e}n wings are observed during very special space weather conditions with extremely low SW densities, that is to say, the simulations performed do not represent the usual conditions for the formation of the Alfv\'{e}n wings for the case of the Earth. Nevertheless, the study provides a generalization of the space weather conditions for the formation of the Alfv\'en wings in exoplanets with an Earth-like magnetosphere. The fit exponents of the simulations with $M_{a} > 1$ for Northward, Southward and Ecliptic IMF orientations are smaller regarding the theoretical value because the effect of the SW dynamic pressure and magnetosphere thermal pressure cannot be neglected. If $|B_{IMF}|$ increases the magnetosphere thermal pressure decreases because the BS plasma is depleted faster as the reconnection regions are located closer to the Earth surface. Consequently, the pressure balance of the simulations with $|B_{IMF}| \ge 20$ nT are dominated by the SW dynamic pressure and the combined effect of the magnetosphere thermal pressure and the Earth magnetic field pressure. Likewise, if $|B_{IMF}| > 20$ nT, the combination of the SW dynamic pressure and the IMF magnetic pressure is mainly balanced by the Earth magnetic field pressure. The Earth-Sun and Sun-Earth IMF orientations show a weak dependency regarding $|B_{IMF}|$, consequence of the magnetosphere bending induced by the IMF in conjunction with the thermal pressure of the magnetosphere, almost unchanged as $|B|_{IMF}$ increases because the BS plasma depletion is rather weak due to the location of the reconnection region above $12 R_{E}$. This results in a magnetopause stand off distance that is nearly constant. In should be noted that the ecliptic clockwise and counter clockwise orientations lead to the same result.

\begin{figure}[h]
\centering
\resizebox{\hsize}{!}{\includegraphics[width=\columnwidth]{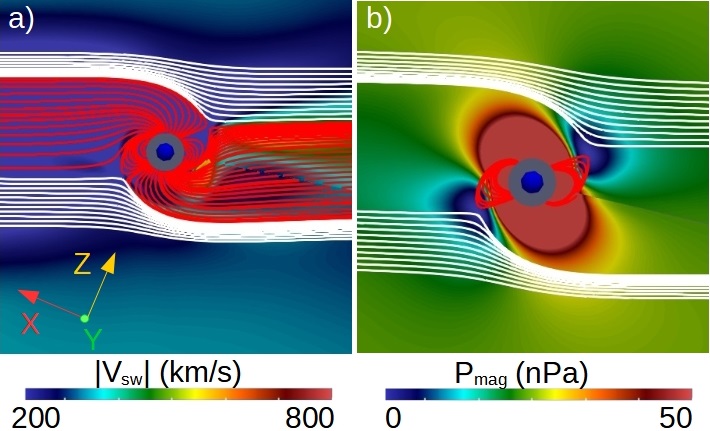}}
\caption{Polar cut (XY plane) of the (a) plasma velocity module (color scale) and (b) magnetic pressure. The red lines indicate the magnetic field lines connected to the Earth surface (red lines) and the white lines the non reconnect IMF lines.}
\label{7}
\end{figure}  

We are now considering the SW effect on the magnetosphere topology. To this end, IMF parameters are kept fixed (Sun-Earth IMF orientation with $|B| = 10$ nT). The IMF intensity in the simulations is small minimizing the IMF effect on the magnetosphere topology. Figure \ref{8} shows $R_{sd}/R_{E}$ for different SW densities (fixed $T_{sw} = 1.8 \cdot 10^{5}$ K and $|v| = 350$ km/s, panel a), SW velocities (fixed $T_{sw} = 1.8 \cdot 10^{5}$ K and $n = 12$ cm$^{-3}$, panel b) and corresponding dynamic pressures (panel c). 

\begin{figure}[h]
\centering
\includegraphics[width=6cm]{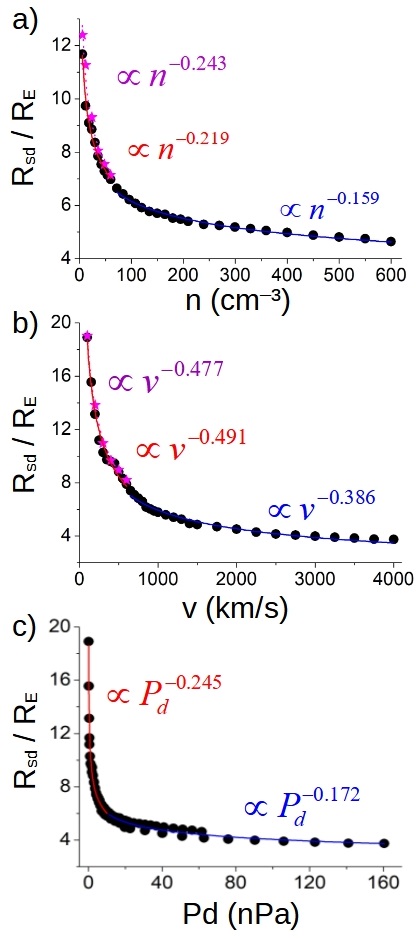}
\caption{Magnetopause stand off distance with respect to (a) the SW density (fixed $v = 350$ km/s), (b) SW velocities (fixed $12 \cdot$ cm$^{-3}$) and (c) dynamic pressure. Sun-Earth IMF orientation with $|B| = 10$ nT. The pink stars indicate the magnetopause stand off distance if $|B| = 0$ nT. The dashed lines indicate the data fit to the expression $R_{sd}/R_{E} = A n^{\alpha}$, $R_{sd}/R_{E} = A |v|^{\alpha}$ and $R_{sd}/R_{E} = A P_{d}^{\alpha}$, respectively. The solid red line indicates the fit line for the data set with $n \le 60$ cm$^{-3}$ and $|v| \le 600$ km/s. The solid blue line indicates the fit line for the data set with $n > 60$ cm$^{-3}$ and $|v| > 600$ km/s. The solid pink line indicates the fit line for the data set with $n \le 60$ cm$^{-3}$ and $|v| \le 600$ km/s and no IMF.}
\label{8}
\end{figure}   

$R_{sd}/R_{E}$ decreases as the SW density or velocity increases, that is to say, a larger dynamic pressure leads to a stronger compression of the BS. It should be noted that, even if the dynamic pressure increases up to $160$ nPa, extreme space weather conditions comparable to a super-ICME, $R_{sd}/R_{E} > 4.5$. Consequently, the direct deposition of the SW toward the Earth surface requires a large distortion of the magnetosphere by the IMF in addition to the BS compression caused by the SW dynamic pressure. Again, the data is fitted to the functions $R_{sd}/R_{E} = A n^{\alpha}$, $R_{sd}/R_{E} = A |v|^{\alpha}$ and $R_{sd}/R_{E} = A P_{d}^{\alpha}$, respectively. In addition, three different data set are used in the regression, the full range of values for the SW density and velocity (dashed black line), $P_{d} < 10$ nPa cases with $n \leq 60$ cm$^{3}$ and $|v| \leq 600$ km/s (red solid line) and $P_{d} > 10$ nPa cases with $n > 60$ cm$^{3}$ and $|v| > 600$ km/s (blue solid line). It should be noted that no plateau is observed in the figures because the minimum dynamic pressure of the simulations is large enough to induce relatively intense deformation of the magnetosphere. Table \ref{3} shows the fitting result.

\begin{table}[h]
\centering
 \begin{tabular}{c | c c | c c}
 & $P_{d} < 10$ & & $P_{d} > 10$ & \\
 & (nPa) & & (nPa) & \\
 SW parameter & A & $\alpha$ & A & $\alpha$ \\ \hline
Density & $17.2$ & $-0.219$ & $12.8$ & $-0.159$ \\
 & $\pm 0.4$ & $\pm 0.007$ & $\pm 0.3$ & $\pm 0.004$ \\
Velocity & $179$ & $-0.491$ & $85$ & $-0.386$ \\
 & $\pm 18$ & $\pm 0.019$ & $\pm 7$ & $\pm 0.014$ \\
Dynamic pressure & $10.60$ & $-0.245$ & $9.0$ & $-0.172$ \\
 & $\pm 0.06$ & $\pm 0.004$ & $\pm 0.3$ & $\pm 0.011$ \\  \hline
\end{tabular}
\caption{Fit parameters of the regressions $R_{sd}/R_{E} = A n^{\alpha}$ (first row), $R_{sd}/R_{E} = A |v|^{\alpha}$ (second row) and $R_{sd}/R_{E} = A P_{d}^{\alpha}$ (third row) for the simulations with low (second and third columns) and large (fourth and fifth columns) $P_{d}$. The standard errors of the regression parameters are included.}
\label{3}
\end{table}

From equation (6), we deduce that the theoretical $\alpha$ exponent is $-0.17$ for the SW density and $-0.33$ for the SW velocity, assuming a negligible effect of the IMF magnetic pressure, SW thermal pressure and magnetosphere thermal pressure in the pressure balance. The fit exponents are close to the theoretical exponents once the SW dynamic pressure is large enough ($P_{d} \ge 10$ nPa) to induce a significant compression of the magnetosphere ($R_{sd}/R_{E} < 7$), thus the pressure balance is dominated by the SW dynamic pressure and the magnetic pressure of the Earth magnetosphere, see solid blue line in figure \ref{8} panels a, b and c. On the other hand, the regression exponents are $25 \%$ larger in the simulations with $P_{d} < 10$ nPa, red solid lines in panels a, b and c. The deviation is caused by the effect of the magnetosphere thermal pressure in the pressure balance. The ratio between the magnetosphere thermal pressure and the SW dynamic pressure increases from $0.2$ to $0.5$ if the SW density decreases from $60$ to $6$ cm$^{-3}$ and from $0.4$ to $0.8$ if the SW velocity decreases from $600$ to $100$ km/s. Consequently the magnetosphere thermal pressure must be included in the pressure balance to calculate correctly the magnetopause stand off distance if the SW dynamic pressure is small. The simulations without IMF (pink stars) and the data fit (pink solid line) indicate the small effect of the Sun-Earth IMF with $|B|_{IMF} = 10$ nT in the pressure balance and the Earth magnetic field topology. The regressions extrapolation indicate a critical $P_{d} \approx 3.5 \cdot 10^{5}$ nPa for the direct deposition of the SW toward the Earth surface, two order of magnitude larger with respect to the $P_{d}$ values during super-ICME for the case of the Earth. Consequently, the direct precipitation of the SW for a relatively weak $|B|_{IMF}$ is extremely unlikely.

Once the effect of the SW on the magnetopause stand off distance is assessed, the following study is dedicated to the effect of the plasma temperature and dynamic pressure on the thickness of the BS ($L_{bs}/R_{E}$), with $L_{bs}$ the distance between the Bow Shock nose and the magnetopause stand off distance at the ecliptic plane in the day side $0^{o}$ longitude. An increase of the SW temperature leads to an increase of the sound speed and the thickness of the BS. On the other hand, a higher dynamic pressure leads to a compression of the BS. Figure \ref{9} shows the $L_{bs}/R_{E}$ values calculated in simulations performed for a range of the SW temperatures (fixed $P_{d} \approx 2$ nPa, panel a) and the $P_{d}$ values (fixed $T_{sw} = 1.8 \cdot 10^{5}$ K, panel b). In addition, the data is fitted to the functions $L_{bs}/R_{E} = A T_{sw}^{\alpha}$ and $L_{bs}/R_{E} = A P_{d}^{\alpha}$, respectively.

\begin{figure}[h]
\centering
\resizebox{\hsize}{!}{\includegraphics[width=\columnwidth]{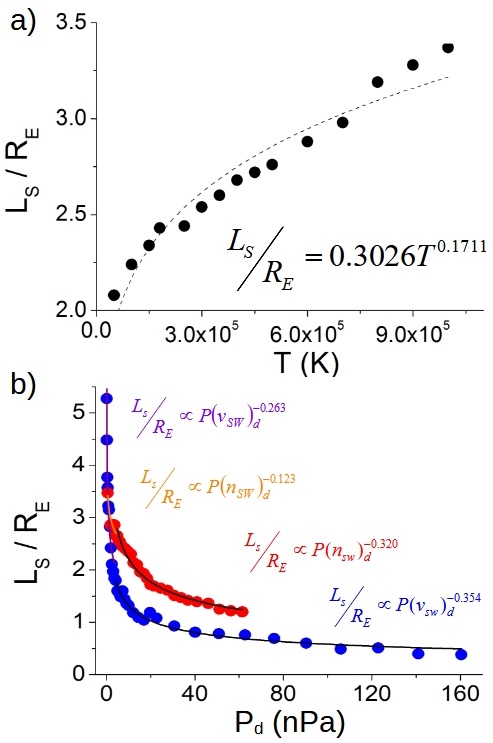}}
\caption{Bow shock width for (a) different SW temperatures (fixed $P_{d} = 1.2$ nPa) and (b) different $P_{d}$ values (fixed $T_{sw} = 1.8 \cdot 10^{5}$ K) if the SW density increases fixed the SW velocity ($350$ km/s, red dots) or the SW velocities increases fixed the SW density ($12 \cdot$ cm$^{-3}$, blue dots). Sun-Earth IMF orientation with $|B| = 10$ nT. The dashed lines indicate the data fit to the expression $ L_{bs}/R_{E} = A T_{sw}^{\alpha} $. The solid black lines indicate the regression $L_{bs}/R_{E} = A P(n_{SW})_{d}^{\alpha}$ in the simulations with the SW velocity fixed and $P_{d} > 4$ nPa and $L_{bs}/R_{E} = A P(v_{SW})_{d}^{\alpha}$ in the simulations with the SW density fixed and $P_{d} < 4$ nPa. The solid violet (orange) line indicates the regression $L_{bs}/R_{E} = A P(n_{SW})_{d}^{\alpha}$ ($L_{bs}/R_{E} = A P(v_{SW})_{d}^{\alpha}$) if $P_{d} < 4$ nPa.}
\label{9}
\end{figure}   

The simulations indicate an increase of the BS width of $\approx 0.4 R_{E}$ if the SW temperature raises from $5 \cdot 10^{4}$ to $2 \cdot 10^{5}$ K (panel a). On the other hand, the BS width decreases $\approx 2.8 R_{E}$ if $P_{d}$ raises from $0.2$ to $160$ nPa (panel b). That is to say, the BS compression caused by the SW $P_{d}$ is around $6-7$ times larger with respect to the BS expansion due to the SW temperature. Also, the BS compression is smaller in the simulations with fixed SW velocity because the plasma temperature and sound speed inside the BS is higher as well as $P_{th}$. In addition, the simulations with $P_{d} < 4$ nPa show a weaker dependence between the BS width and $P_{d}$ (please compare the regression parameters of the simulations with $P_{d} > 4$ nPa and $P_{d} < 4$ nPa), because the magnetosphere thermal pressure is comparable to $P_{d}$. By contrast, as the simulation $P_{d}$ increases and the role of the magnetosphere thermal pressure is less important in the pressure balance, the dependence between the BS width and $P_{d}$ increases. It should be mentioned that, the range of SW temperature and $P_{d}$ values highlighted includes the typical SW parameters during regular and extreme space weather conditions \citep{Cliver,Mays}.

In summary, the magnetopause stand off distance calculated in the simulations reveals the key role of the IMF and SW on the distortion of the Earth magnetosphere for regular and extreme space weather conditions. The data regressions show clear differences in the pressure balance for super-Alfv\'{e}nic and sub-Alfv\'{e}nic configurations, as well as the important role of the magnetosphere thermal pressure in the determination of the magnetopause stand off distance if the SW dynamic pressure and IMF magnetic pressure are low. It should be noted that the range of magnetopause stand off distance calculated is comparable to the results obtained by other authors \citep{Song,Kabin,Lavraud,Ridley,Meng,Wang5}. Present study contribution entails a larger sample of space weather configurations thanks to the extended parametric studies performed, as well as the detail analysis of the topological deformation trends linked to the SW and IMF properties.

\subsection{Reconnection region tracking for different IMF orientations and intensities}

\begin{figure}[h]
\centering
\resizebox{\hsize}{!}{\includegraphics[width=\columnwidth]{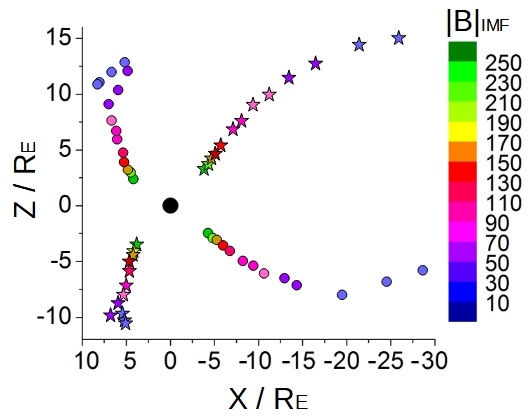}}
\caption{Location of the reconnection regions in the XY plane if $|B_{IMF}|$ increases from $10$ to $250$ nT for Earth-Sun and Sun-Earth IMF orientations. $P_{d} = 1.2$ nPa and $T_{sw} = 1.8 \cdot 10^{5}$ K. Symbol color indicates the $|B_{IMF}|$ value. The stars indicate the reconnection region for the Sun-Earth IMF orientation. The circles indicate the reconnection region for the Earth-Sun IMF orientation.}
\label{10}
\end{figure}

\begin{figure}[h]
\centering
\resizebox{\hsize}{!}{\includegraphics[width=\columnwidth]{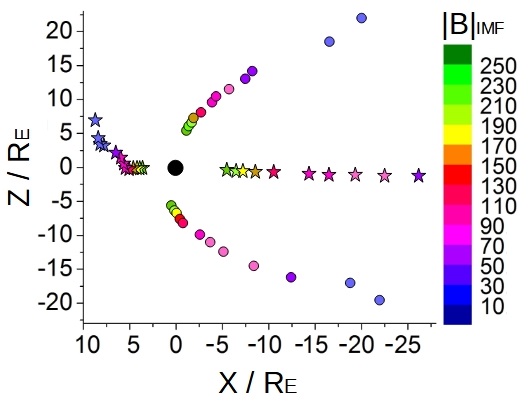}}
\caption{Location of the reconnection regions in the XY plane if $|B_{IMF}|$ increases from $10$ to $250$ nT for Northward and Southward IMF orientations. $P_{d} = 1.2$ nPa and $T_{sw} = 1.8 \cdot 10^{5}$ K. Color contour indicates the $|B_{IMF}|$ value. The yellow (gray) star indicates the reconnection region at the day (night) side for a Southward IMF orientation. The yellow (gray) circle indicates the reconnection region near the North (South) pole for the Northward IMF orientation.}
\label{11}
\end{figure}

This section is dedicated to track the location of the reconnection regions for different IMF orientation and intensities. Figure \ref{10} indicates the location of the reconnection regions in the XY plane for Sun-Earth and Earth-Sun IMF orientations as $|B_{IMF}|$ increases from $10$ to $250$ nT. Likewise, figure \ref{11} shows the same study for Northward and Southward IMF orientations. The reconnection in the simulations is identified as the region where the magnetic field intensity goes to zero.

The reconnection region for the Sun-Earth IMF orientation at the day side moves towards the South pole as $|B_{IMF}|$ increases, showing a large Northward displacement although smaller in the Earth-ward direction. On the other hand, the reconnection at the night side moves towards the North pole and the larger displacement is done in the Sun-ward direction with respect to the Southward displacement. Regarding the Earth-Sun IMF orientation, the reconnection in the day side moves Southward towards the North pole although the reconnection in the night side moves toward the South pole. The differences between the Sun-Earth and Earth-Sun orientations are caused by the North-South bending of the Earth magnetosphere. It should be noted that the reconnection region at the night side is located outside the computational domain for the simulations with $|B_{IMF}| < 30$ nT, thus this data is not included in the analysis.

The reconnection region in the simulations with Southward IMF orientation are located closer to the equatorial plane and the Earth surface as $|B_{IMF}|$ increases. In the day side, the reconnection displaces Southward and Earth-ward. Regarding the Northward IMF orientation, the reconnections are located closer to the poles as $|B_{IMF}|$ increases .The reconnection region at the night side is outside the computational domain for the simulations with Southward IMF and $|B_{IMF}| < 60$ nT as well as for the simulations with Northward IMF and $|B_{IMF}| < 40$ nT and are not considered.

\begin{figure*}[h]
\centering
\resizebox{\hsize}{!}{\includegraphics[width=\columnwidth]{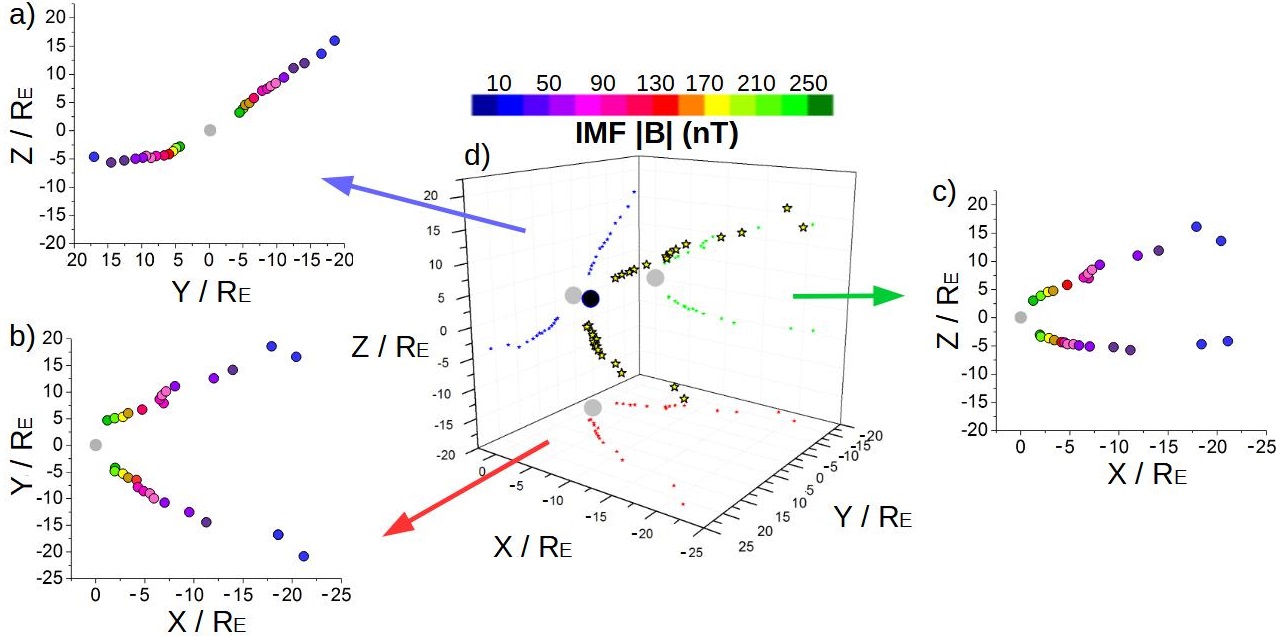}}
\caption{Location of the reconnection region if $|B_{IMF}|$ increases from $10$ to $250$ nT for a ecliptic ctr-clockwise IMF orientation. $P_{d} = 1.2$ nPa and $T = 1.8 \cdot 10^{5}$ K. Color contour indicates the $|B_{IMF}|$ value. Panel (a) indicates the projection in the YZ plane, (b) the projection in the XZ plane and (c) the projection in the XY plane. Panel (d) shows the 3D view.}
\label{12}
\end{figure*}

Figure \ref{12} shows the location of the reconnection region for an Ecliptic ctr-clockwise IMF orientation as $|B_{IMF}|$ increases. The clockwise case is not included because the Earth magnetosphere shows a symmetric topology deformation with respect to the Ecliptic IMF orientations. The analysis is more complex regarding the other IMF orientations because the reconnections are not located in the XY plane and should be tracked in 3D. 

The reconnection regions move towards the planet surface as the $|B_{IMF}|$ increases following the East/West tilt induced in the Earth magnetosphere. It should be noted that the reconnection region is outside the computational domain for the simulations with $|B_{IMF}| < 20$ nT.

Summarizing, the location of the reconnection regions is critical to understand the effect of the IMF orientation and intensity on the Earth magnetosphere topology. The study reveals a large variation of the Earth magnetosphere topology in the range of IMF intensities and orientations analyzed. Consequently, the SW injection into the inner magnetosphere and the plasma flows towards the Earth surface are very different regarding the IMF configuration. Table \ref{4} shows the $\%$ of the reconnection displacement for different IMF orientations (defined as  $100 \cdot \Delta r_{max} / \Delta r_{min}$ with $\Delta r = \sqrt{\Delta x^2 + \Delta y^2 +\Delta z^2}$) between the simulations with $|B_{IMF}|=250$ nT and $|B_{IMF}|_{min}$ (third and fifth columns). The third and fifth columns also indicate if the reconnection regions are located in the day side, night side, North pole, South pole, West or East of the magnetosphere. If the reconnection region is located outside the computational domain (because $|B_{IMF}|$ is small), the simulation is not included in the analysis and the configuration with the lowest $|B_{IMF}|$ with the reconnection region inside the computational domain is indicated in the table (second and fourth columns).

\begin{table}[h]
\centering
\begin{tabular}{c | c c c c}
IMF & $|B_{IMF}|_{min}$ & DS & $|B_{IMF}|_{min}$ & NS \\ 
 & (nT) &  ($\%$)  & (nT) & ($\%$) \\ \hline
Sun-Earth & $10$ & $46.37$ & $30$ & $17.02$\\
Earth-Sun & $10$ & $35.58$ & $30$ & $16.97$\\
Southward & $10$ & $34.42$ & $60$ & $20.93$\\ \hline 
 &  & North P. & & South P. \\ 
Northward & $40$ & $19.39$ & $40$ & $18.94$\\ \hline 
 &  & West & & East \\ 
Ctr-cw Ecliptic & $20$ & $18.54$ & $20$ & $19.30$\\ \hline
\end{tabular}
\caption{IMF orientation (first column). IMF intensity of the simulation with the lowest $|B_{IMF}|$ and the reconnection region located inside the computational domain (second and fourth columns). Maximal reconnection displacement $\%$ between the simulations with maximum and minimum $|B_{IMF}|$ at the day side (night side) for Sun-Earth, Earth-Sun and Southward IMF orientations, North (South) pole for the Northward orientation and the West (East) of the magnetosphere for the ctr-cw ecliptic IMF orientation (third and fifth columns, respectively).}
\label{4}
\end{table}

\subsection{Open-close field line boundary for different IMF orientations and intensities}

\begin{figure}[h]
\centering
\resizebox{\hsize}{!}{\includegraphics[width=\linewidth]{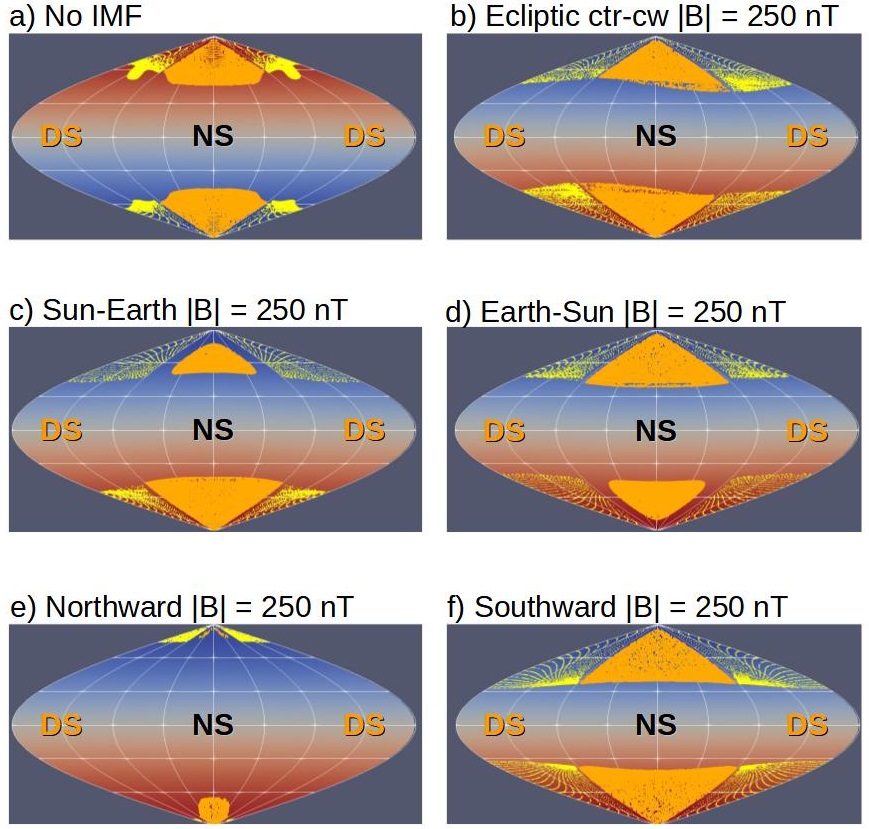}}
\caption{Sinusoidal (Sanson-Flamsteed) projection of the open-close magnetic field line boundary at $R / R_{E} = 2.05$ for (a) $|B_{IMF}| = 0$ nT, (b) Ecliptic ctr-clockwise $|B_{IMF}| = 250$ nT, (c) Sun-Earth $|B_{IMF}| = 250$ nT, (d) Earth-Sun $|B_{IMF}| = 250$ nT, (e) Northward $|B_{IMF}| = 250$ nT and (f) Southward $|B_{IMF}| = 250$ nT. Fixed $P_{d} = 1.2$ nPa and $T_{sw} = 1.8 \cdot 10^{5}$ K. The yellow (orange) dots indicates the open magnetic field lines at the Earth day (night) side.}
\label{13}
\end{figure}

The modification of the Earth magnetosphere topology by the IMF also modifies the ratio between open / close magnetic field lines at the Earth surface. The Earth surface covered by open field lines is more vulnerable to extreme SW conditions because the plasma precipitate along the magnetic field lines toward the surface. If a large amount of SW is injected in the inner magnetosphere through the reconnection regions, the planet surface covered by open field lines suffers an enhancement of the plasma flows. Consequently, it is important to study how the IMF intensity and orientation modifies the latitude of the open-close field line boundary (OCB). Figure \ref{13} shows the open magnetic field lines at $R / R_{E} = 2.05$ for different IMF intensities and orientations (fixed $P_{d} = 1.2$ nPa and $T_{sw} = 1.8 \cdot 10^{5}$ K). The OCB lines are identified by an iterative method that calculates the last close magnetic field line connecting the inner boundary of the computational domain and concentric spheres with radius between $R_{sd}/R_{E}$ and $(R_{sd}+R_{E})/R_{E}$. The magnetotail is fully located inside the simulation domain in the configurations analyzed, please see the Appendix C for further discussion.

The increase of $|B_{IMF}|$ for a Sun-Earth IMF orientation (panel a) causes a decrease of the OCB latitude at the day side, particularly large at the North Hemisphere. This results is consistent with the Southward bending of the Earth magnetosphere, promoting a stronger erosion of the Earth magnetic field by the IMF in the North Hemisphere. The East/West tilt of the magnetosphere caused by IMF orientations in the Ecliptic plane (panel b) is also observed in the open field line distribution, leading to a large longitudinal and latitudinal OCB dependency. Regarding the Sun-Earth and Earth-Sun IMF orientations (panels c and d), the OCB is asymmetric with respect to the day and night sides. On the other hand, the displacement of the reconnection regions toward the Earth magnetic axis (equatorial plane) for a Northward (Southward) IMF orientation leads to a displacement of the OCB towards a higher (lower) latitude (panels e and f). It should be noted that the Southward IMF orientation leads to the lowest OCB latitude at the day and night side for both Hemispheres. 

\begin{figure}[h]
\centering
\resizebox{\hsize}{!}{\includegraphics[width=\columnwidth]{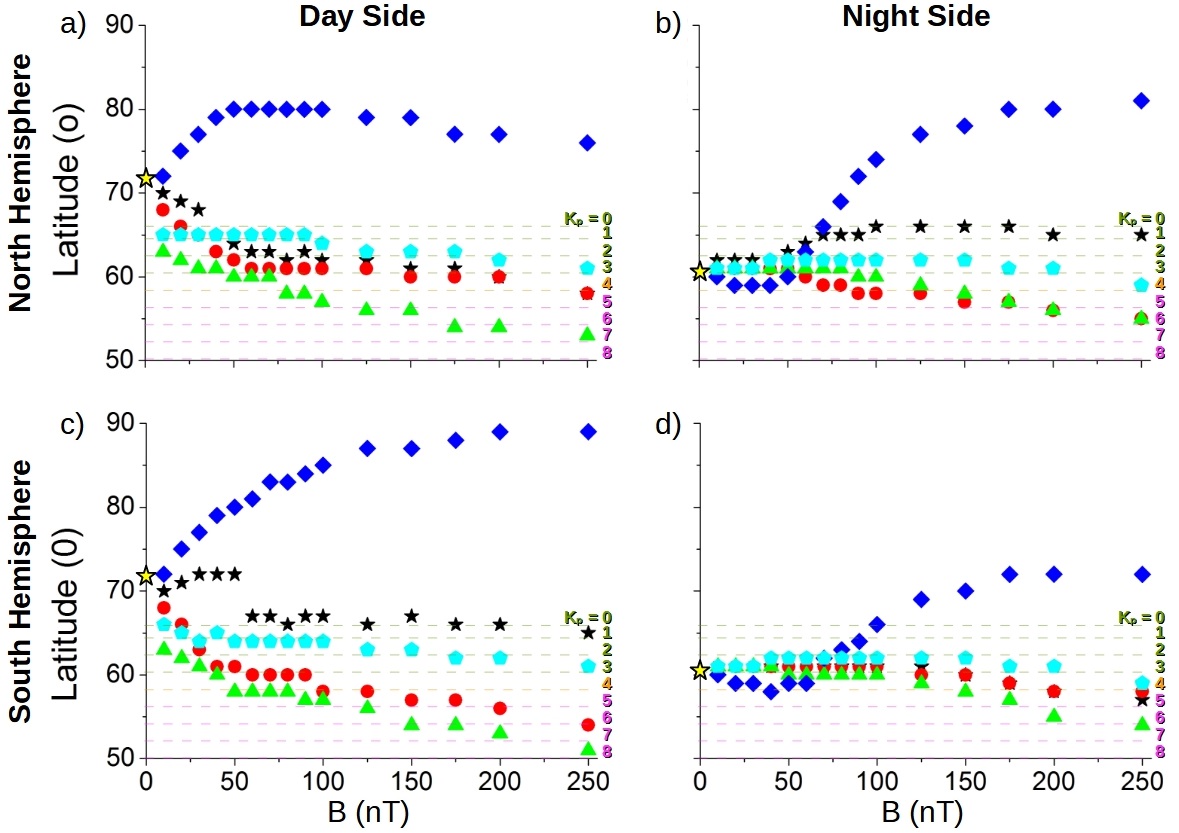}}
\caption{OCB latitude with respect to the IMF orientation and $|B_{IMF}|$ calculated at the North Hemisphere (a) day side ($0^{o}$ longitude) and (b) night side ($180^{o}$ longitude), South Hemisphere (c) day side and (b) night side. Fixed $P_{d} = 1.2$ nPa and $T_{sw} = 1.8 \cdot 10^{5}$ K. The dashed horizontal lines indicate the $K_{p}$ index. The yellow star indicates the OCB latitude if $|B_{IMF}| = 0$.}
\label{14}
\end{figure}

The latitude of the OCB at the Earth surface is calculated mapping the magnetic field lines obtained in the simulations with the magnetic field of a dipole without the distortion of the SW and IMF. The magnetic field line mapping is described in the Appendix B, where it is shown that below $2R_{E}$ there is a good agreement between the unperturbed and perturbed dipole, thus the OCB line latitude at the Earth surface can be extrapolated with reasonable confidence. Figure \ref{14} indicates the OCB latitude with respect to the IMF orientation and $|B_{IMF}|$ calculated in the Earth day side ($0^{o}$ longitude) and night side ($180^{o}$ longitude) at the North and South Hemispheres. In addition, the OCB latitude is compared with the latitude of the auroral oval associated with different $K_{p}$ index. The $K_{p}$ index indicates the global geomagnetic activity, taking values from $0$ for the case of weak geomagnetic activity to $9$ if there is an extreme geomagnetic activity \citep{Menvielle,Thomsen}.

The OCB latitude at the North Hemisphere day side (panel a) decreases from $70^{o}$ to $58^{o}$ as $|B_{IMF}|$ increases for the Sun-Earth IMF orientation. Regarding the Earth-Sun IMF orientation, the range of OCB latitudes is slightly smaller, between $68 - 58^{o}$, due to the Northward bending of the magnetosphere at the day side, leading to slightly differences with respect to the Sun-Earth IMF at the North Hemisphere although larger differences at the South Hemisphere (panel c). On the other hand, the Northward IMF orientation leads to an increase of the OCB latitude at the North Hemisphere up to $80^{o}$ if $|B_{IMF}| = 100$ nT, decreasing to $76^{o}$ if $|B_{IMF}| = 250$ nT due to the combined effect of the magnetosphere compression and the tilt. On the other hand, the OCB latitude at the South Hemisphere increases with $|B_{IMF}|$ reaching $88^{o}$ for $|B_{IMF}| = 250$ nT simulation. It should be noted that the trend of the OCB latitude regarding $|B_{IMF}|$ is inverted between Hemispheres at the planet night side (see fig \ref{10}), again due to the effect of the magnetic field tilt. For the Southward IMF orientation, the OCB latitude decreases as $|B_{IMF}|$ increases, between $63-53^{o}$ in the North Hemisphere and $63-51^{o}$ in the South Hemisphere if the simulations with $|B_{IMF}| = 10$ and $250$ nT are compared due to the Earth magnetic field erosion at the equatorial region. For the Ecliptic ctr-clockwise orientation, the OCB latitude slightly decreases as $|B_{IMF}|$ increases, from $65^{o}$ to $61^{o}$ if $|B_{IMF}|$ increases from $10$ to $250$ nT, because the West/East asymmetry induced in the magnetosphere has a lesser effect on the OCB latitude. The latitude of the auroral oval for different $K_{p}$ index is included in the panels and compared with the OCB latitude at the Earth day and night side providing an approximation of the $K_{p}$ index in the simulations. The largest variation of the OCB line with respect to the $K_{p}$ index as $|B_{IMF}|$ enhances is observed for the Earth-Sun and Southward IMF orientations. Consequently, Earth-Sun and Southward IMF orientations can lead to large geomagnetic activities, result consistent with previous studies by \citet{Schatten,Boroyev}. The Sun-Earth, Northward and Ecliptic IMF orientations lead to $K_{p} \le 4$ if $|B_{IMF}| = 250$ nT, thus the geomagnetic activity caused is relatively quiet. It should be noted that the model latitudinal resolution is $4^{o}$, thus the uncertainly on the $K_{p}$ index prediction is $\pm 2$, enough to distinguish between quiet ($K_{p} < 3$), moderate ($3 \le K_{p} \le 6$) and strong ($ K_{p} > 6$) auroral activity.

\begin{figure}[h]
\centering
\resizebox{\hsize}{!}{\includegraphics[width=\columnwidth]{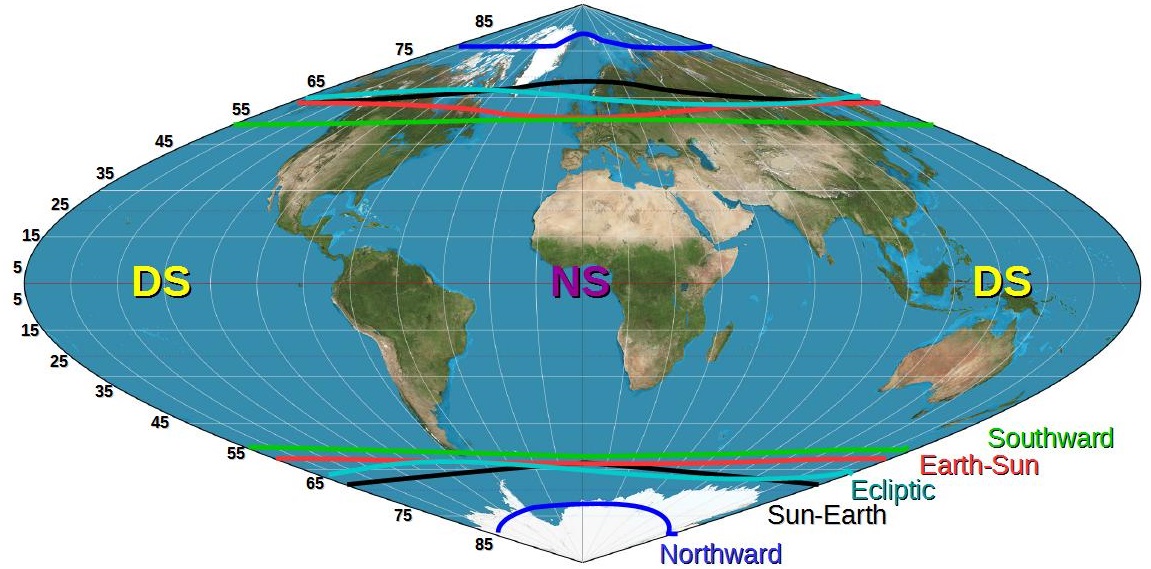}}
\caption{Schematic view of the OCB line for Sun-Earth (black line), Earth-Sun (red line), Northward (blue line), Southward (green line) and Ecliptic ctr-clockwise (cyan lines) with $|B_{IMF}| = 250$ nT . Fixed $P_{d} = 1.2$ nT and $T_{sw} = 1.8 \cdot 10^{5}$ K. The location of the DS/NS with respect to the continents is irrelevant.}
\label{15}
\end{figure}

In summary, the OCB latitude, and so the exposition of the Earth surface to the plasma flows from the magnetosheath, shows a clear dependency with respect to the space weather conditions, leading to a large decrease of the OCB latitude particularly for an intense Southward oriented IMF. For example, fig \ref{15} shows the OCB line for different IMF orientations with $|B_{IMF}| = 250$ nT, indicating that the South of Canada and the North of England are exposed if the IMF orientation is Southward, thus the electric grid of these countries are endanger.  It should be noted that similar trends were obtained by other authors with respect to the IMF orientation and intensity \citep{Lopez,Kabin,Wild,Wang7,Burrell}, results extended and refined in the present paper thanks to the large sample of parametric studies performed, including a forecast of the $K_{p}$ index variation with respect to the IMF module and orientation.

\subsection{Combined effect of the dynamic pressure and IMF orientation/intensity on the Earth magnetic field topology}

\begin{figure}[h]
\centering
\resizebox{\hsize}{!}{\includegraphics[width=\columnwidth]{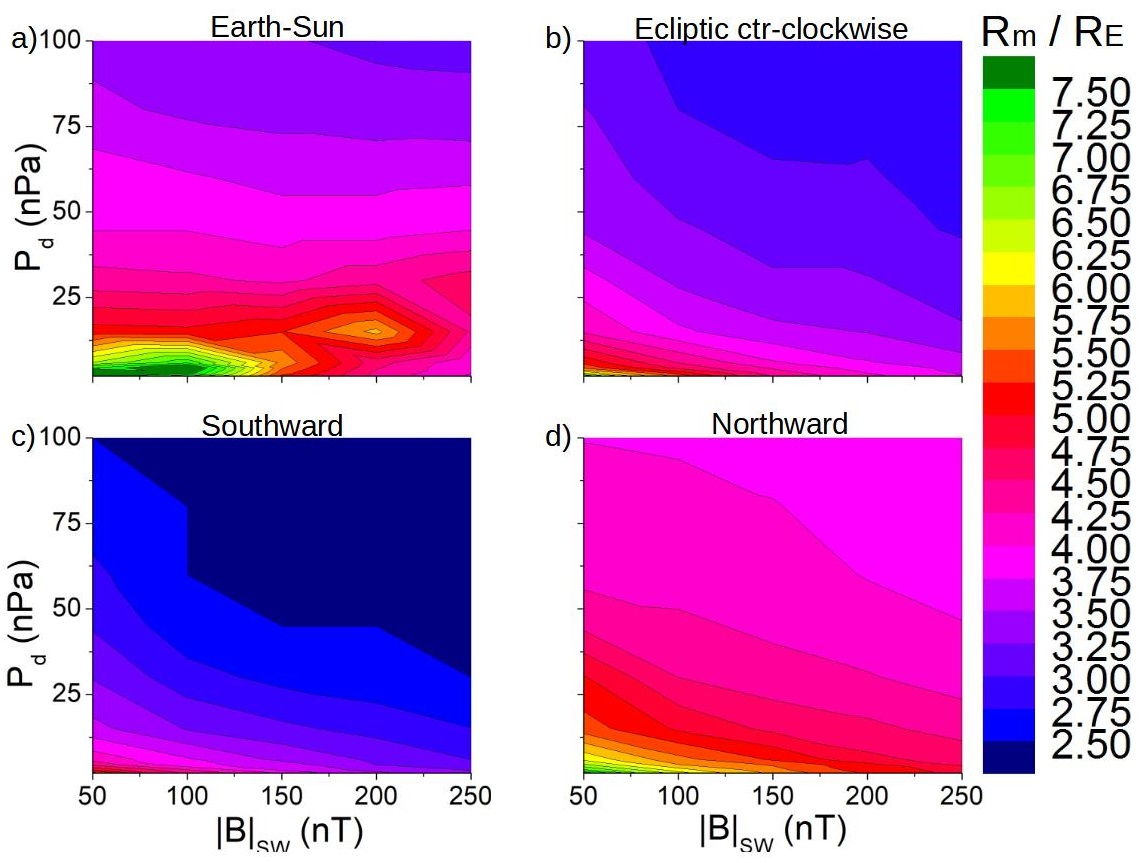}}
\caption{Iso-contour of the magnetopause stand off distance for different $P_{d}$ and $|B_{IMF}|$ values if the IMF is oriented (a) Earth Sun, (b) Ecliptic ctr-clockwise, (c) Southward and (d) Northward. Fixed $T = 1.8 \cdot 10^{5}$ K.}
\label{16}
\end{figure}

A complete parametric study of the Earth magnetosphere topology with respect to the space weather conditions requires the combined effect of the SW dynamic pressure and the IMF module and orientation. On that aim, figure \ref{16} indicates the magnetosphere stand off distance with respect to the IMF orientation and module (for $|B_{IMF}| = 50, 100, 150, 200$ and $250$ nT) and the SW dynamic pressure ($P_{d} = 1.2, 1.5, 3, 4.5, 6, 15, 30, 45, 60, 80$ and $100$ nPa). The range of parameters include regular and extreme space weather conditions (ICME). Ecliptic clockwise and counter-clockwise IMF orientations lead to the same results, thus only the counter-clockwise case is analyzed. Regarding the Sun-Earth orientation, this case is not included because, in spite of the magnetic field tilt, there is an North-South symmetry with respect of the Earth-Sun orientation so the simulations results are similar.

The IMF orientation that leads to the smallest magnetopause stand off distance distance with respect to $|B_{IMF}|$ and $P_{d}$ is the Southward orientation. The simulations indicate that, for $P_{d}$ and $|B|_{IMF}$ values consistent with extreme space conditions as a ICMEs impacting the Earth ($|B|_{IMF} \approx 100$ nT and $P_{d} \approx 30$ nPa), there is no direct precipitation of the SW towards the Earth surface. The smallest $R_{sd}/R_{E} = 2.92$ is obtained for the Southward IMF orientation. In addition, super-CMEs with $|B|_{IMF} = 250$ nT and $P_{d} = 100$ nPa and a IMF oriented in the Southward direct just lead to a  $R_{sd}/R_{E}$ slightly below $2.5$, above $2.8$ for the rest of IMF orientations. The balance between the dynamic pressure and IMF intensity is particularly complex for the Earth-Sun IMF orientation for simulations with $P_{d} < 30$ nPa, leading to an increase of $R_{sd}/R_{E}$ as $|B|_{IMF}$ increases caused by the North-South deformation induced at the day/night side of the magnetosphere, respectively. On the other hand, the simulations with $P_{d} > 30$ nPa, $R_{sd}/R_{E}$ show a weak dependency with respect to $|B|_{IMF}$ because the compression of the BS is partially counter-balanced by the North-South asymmetry induced in the magnetosphere.

\begin{figure}[h]
\centering
\resizebox{\hsize}{!}{\includegraphics[width=\columnwidth]{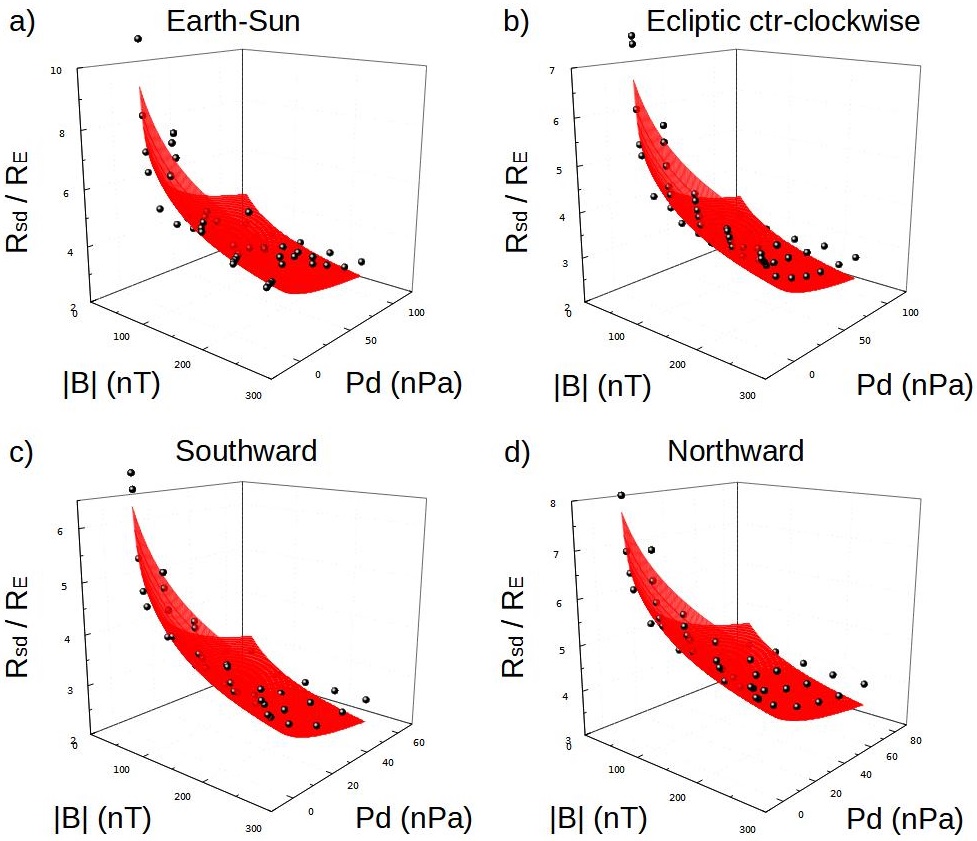}}
\caption{Plot of the surface function $R_{sd} / R_{E} = A |B|_{IMF}^{\alpha} P_{d}^{\beta}$ if the IMF is oriented in the (a) Earth Sun, (b) ecliptic ctr-clockwise, (c) Southward and (d) Northward directions. Fixed $T_{sw} = 1.8 \cdot 10^{5}$ K.}
\label{17}
\end{figure}

Next, $R_{sd}/R_{E}$ data is fitted with respect to $P_{d}$ and $|B|_{IMF}$ by the surface function $R_{sd} / R_{E} = A |B|_{IMF}^{\alpha} P_{d}^{\beta}$. The regression results are indicated in the table \ref{5} and figure \ref{17}. The regression analysis for the Southward IMF orientation only includes $P_{d} < 60$ nPa values, thus the simulations with $R_{sd} / R_{E} < 2.5$ are not included in the study.

\begin{table}[h]
\centering
\begin{tabular}{c | c c c }
IMF & A & $\alpha$ & $\beta$ \\ \hline
Earth-Sun & $40$ & $-0.35$ & $-0.16$\\
 & $\pm 8$ & $\pm 0.04$ & $\pm 0.02$\\ 
Northward & $17.2$ & $-0.196$ & $-0.122$\\
 & $\pm 1.3$ & $\pm 0.016$ & $\pm 0.007$\\
Southward & $20.2$ & $-0.286$ & $-0.175$\\
 & $\pm 1.6$ & $\pm 0.016$ & $\pm 0.008$\\
Ecliptic & $19.2$ & $-0.260$ & $-0.143$\\
 & $\pm 1.8$ & $\pm 0.019$ & $\pm 0.008$\\ \hline
\end{tabular}
\caption{Fit parameters of the regression $R_{sd} / R_{E} = A |B|_{IMF}^{\alpha} P_{d}^{\beta}$ and standard errors.}
\label{5}
\end{table}

The expected exponent from the equation (6) are $\alpha = -0.33$ and $\beta = -0.17$, although the regressions show clear deviations mainly caused by the Earth magnetic field reconnection with the IMF, particularly in the simulations with large $|B|_{IMF}$, as well as the pressure generated by the particles inside the BS in the simulations with low $P_{d}$. The largest deviation of the $\alpha$ exponent is observed for the Ecliptic and Northward IMF orientation, because there is a strong West-East tilt and pole-ward stretching induced in the magnetosphere further promoted in simulations with large $P_{d}$, respectively. Regarding the Southward IMF orientation, the $\alpha$ exponent is smaller regarding the theoretical value due to the erosion induced in the Earth magnetic field at the equatorial region. On the other hand, the Earth-Sun IMF orientation shows an $\alpha$ exponent closer to the theoretical value because the induced Northward bending of the magnetosphere is smaller as the simulation $P_{d}$ increases. It should be noted that the $\beta$ exponent of the regressions show a reasonable agreement with the theoretical exponent, although the deviation is significant regarding the Ecliptic and Northward IMF orientation cases for the reasons already mentioned.

In conclusion, the SW dynamic pressure, IMF intensity and orientation are the main parameters required to study the response of the Earth magnetosphere to the space weather conditions. Thus, the trends of the topological deformations identified by the parametric study can be generalized, providing a new tool to analyze the consequences of the magnetosphere distortion, the topic of the following section.

\section{Analysis applications}
\label{Applications}

This section shows several application of the present study conclusions regarding the direct exposition of satellites at different orbits to the SW for different space weather conditions, the Earth habitability along the Sun main sequence and a ICME classification with respect to $|B|_{IMF}$, $P_{d}$ and $Dst$ parameters.

\subsection{Forecast of the space weather conditions for the SW precipitation towards the Earth surface}

\begin{figure}[h]
\centering
\resizebox{\hsize}{!}{\includegraphics[width=\columnwidth]{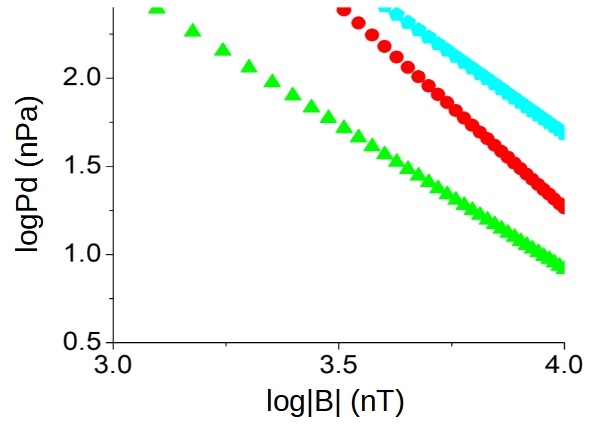}}
\caption{Critical $|B|_{IMF}$ required for the direct precipitation of the SW towards the Earth surface with respect to $P_{d}$ and the IMF orientation. Fixed $T_{sw} = 1.8 \cdot 10^{5}$ K.}
\label{18}
\end{figure}

Figure \ref{18} shows the critical $|B|_{IMF}$ required for the direct precipitation of the SW towards the Earth surface with respect to $P_{d}$ and the IMF orientation. The SW precipitates directly towards the Earth surface if the magnetopause stand off distance of the simulation is the same regarding the Earth radius. Thus, the critical  $|B|_{IMF}$ is calculated from the regression parameters taking $R_{sd} / R_{E} = 1$, thus $ ln (|B|_{IMF,c}) = ln [(A P^{\beta})^{-1}] / \alpha$.

The direct precipitation of the SW requires the combination of extreme $P_{d}$ and $|B|_{IMF}$ values well above the space weather conditions at the Earth even during super-ICME. For example, a Southward IMF orientation with $|B|_{IMF} = 1000$ nT requires $P_{d} \geq 355$ nPa while an Earth-Sun IMF orientation requires $P_{d} \geq 3660$ nPa, $5-4$ times larger with respect to a super-ICME, respectively. It should be noted that Ecliptic and Northward IMF orientations require even larger $|B|_{IMF}$ and $P_{d}$ combinations for the direct precipitation of the SW.

\subsection{Space weather conditions for the direct exposition of satellites to the solar wind}

\begin{figure}[h]
\centering
\includegraphics[width=6cm]{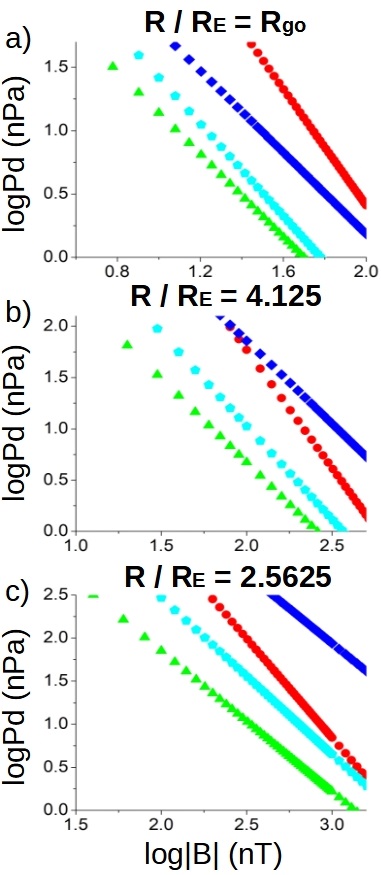}
\caption{Critical $P_{d}$ to reduce the magnetopause stand off distance below (a) the Geostationary orbit (b) Medium orbit at $R_{mo} = R / R_{E} = 4.125$ and (c) Medium orbit at $2.5625$ for different IMF intensities and orientations. Fixed $T = 1.8 \cdot 10^{5}$ K.}
\label{19}
\end{figure}

The direct exposition to the SW can inflict the failure of the satellite electronics by radiation damage and charging. The Earth magnetic field brings protection to the spacecrafts although the distortion of the magnetosphere driven by the space weather conditions can lead to excursions outside the inner magnetosphere along the satellite orbit, particular at the Earth day side where the magnetosphere is compressed by the SW and eroded by the IMF. Consequently, it is important to analyze the space weather conditions that can lead to the direct exposition of satellites at different orbits to the SW. The satellite orbits around the Earth are classified in Low orbits (below $2000$ km), Medium orbits (between $2000 - 35786$ km), Geosynchronous and Geostationary orbits (at $35786$ km) and High orbits (above $35786$ km). Figure \ref{19} indicates the critical $P_{d}$ for different IMF intensities and orientations required to reduce the magnetopause stand off distance below the Geostationary orbit $R_{go} = R / R_{E} \approx 6.6$ (panel a) and Medium orbits at $R_{mo} = R / R_{E} = 4.125$ ($20000$ km, panel b) and $2.5625$ ($10000$ km, panel c).

Regarding the satellites at the geostationary orbit (panel a), during regular space weather conditions with $P_{d} \approx 1$ nPa there is a transit outside the inner magnetosphere at the Earth day side if the IMF orientation is Earth-Sun and $|B|_{IMF} > 150$ nT, decreasing to $130$ nT for a Northward IMF. The $|B|_{IMF}$ decreases to $50 - 60$ nT if the IMF is Southward or Ecliptic ctr-cw. That is to say, Southward and Ecliptic IMF orientations are adverse for geostationary satellites because $R_{sd}/R_{E}$ decrease below $R_{go}$ due to the erosion of the magnetic field at the day side and the East-West asymmetry driven in the magnetosphere, respectively. The same way, if the space weather conditions lead to an enhancement of $P_{d}$, the Geostationary satellites are exposed for Southward IMF with $|B|_{IMF} = 10$ nT and $P_{d} \approx 14$ nPa, as well as Ecliptic IMF with $|B|_{IMF} = 10$ nT and $P_{d} \approx 26$ nPa.

Concerning Earth Medium orbits, a satellite at $20000$ km is exposed during regular space weather conditions with $P_{d} \approx 1$ nPa if $|B|_{IMF} > 260$ nT for a Southward IMF, $|B|_{IMF} > 360$ nT for a Ecliptic IMF, $|B|_{IMF} > 600$ nT for a Earth-Sun IMF and $|B|_{IMF} > 1450$ nT for a Northward IMF. Consequently, satellites at $20000$ km and lower orbits are protected by the magnetosphere during regular space weather conditions because the critical $|B|_{IMF}$ is too large. On the other hand, extreme space weather conditions lead to exposed satellites at $20000$ km for Southward IMF with $|B|_{IMF} = 40$ nT and $P_{d} \approx 20$ nPa, Ecliptic IMF with $|B|_{IMF} = 40$ nT and $P_{d} \approx 55$ nPa, Earth-Sun IMF with $|B|_{IMF} = 100$ nT and $P_{d} \approx 60$ nPa as well as Northward IMF with $|B|_{IMF} = 100$ nT and $P_{d} \approx 70$ nPa. In addition, satellites at $10000$ km orbit are only exposed if the IMF is Southward, $|B|_{IMF} = 100$ nT and $P_{d} \approx 70$ nPa.

\subsection{ICME classification}

Most of the ICMEs that impact the Earth, around $1000$ each Sun cycle, show an averaged plasma velocity smaller than $500$ km/s and IMF intensities below $15$ nT leading to geomagnetic storms with $Dst < -50$ nT \citep{Cane2}. Super ICME events similar to the 'Carrington event' are less frequent, around once each century \citep{Riley2}, although the potential damage in space and ground technological resources is large \citep{Baker4,Eastwood,Eastwood2}. Other examples of super-ICME that did not impact the Earth were analyzed by \citep{Liu2} using STEREO data, indicating plasma velocities around $2000$ km/s, a density of $100$ cm$^{-3}$ ($P_{d} \approx 330$ nPa) and $|B|_{IMF} \approx 100$ nT ($Dst = -600$ to $-1100$ nT). The space weather conditions during ICMEs can also be modeled using ENLIL \citep{Odstrcil2,Odstrcil3}, EUropean Heliospheric FORecasting Information Asset (EUHFORIA) \citep{Pomoell} and Susanoo model \citep{Shiota}, between others. Combining satellite data and modeling results, a classification of the ICMEs is proposed in table \ref{6} with respect to the SW dynamic pressure, IMF intensity and $Dst$ parameter.

\begin{table}[h]
\centering
\begin{tabular}{c | c c c }
ICME type & $P_{d}$ & $|B|_{IMF}$ & $Dst$ \\
 & (nPa) & (nT) & (nT) \\ \hline
Common & $< 40$ & $< 50$ & $> -50$ \\
Strong & $[40 , 100]$ & $[50 , 100]$ & $[-50 , -200]$ \\
Super & $> 100$ & $> 100$ & $< -200$ \\
\end{tabular}
\caption{ICME classification with respect to the SW dynamic pressure, IMF intensity and $Dst$ parameter.}
\label{6}
\end{table}

If the results of the present study are analyzed in the context of the proposed ICME classification, the direct precipitation of the SW toward the Earth surface is very unlikely, even for the case of super-ICME. The direct SW precipitation during a super-ICME for a Southward IMF orientation requires extreme space weather conditions values well above the expected range of $P_{d}$ and $|B|_{IMF}$ values, as it was discussed in previous sections. Nevertheless, the extreme space weather conditions inside the category of Strong ICMEs already leads to magnetopause stand off distances around $R_{sd} / R_{E} \approx 2.5$ if $P_{d} = 60$ nPa and $|B|_{IMF} = 100$ nT for a Southward IMF orientation, $R_{sd} / R_{E} \approx 3.8$ for a Earth-Sun IMF orientation, $R_{sd} / R_{E} \approx 3.2$ for a Ecliptic IMF orientation and $R_{sd} / R_{E} \approx 4.1$ for a Northward IMF orientation. Consequently, strong ICME are a threat to satellites at Geostationary, High and Medium orbits. Particularly, the Geostationary satellites are above the magnetopause for all the range of space weather condition inside the strong ICME category and IMF orientations. In addition, the Medium orbit satellites at $20000$ km are above the magnetopause if the IMF is Southward for all the range of space weather condition inside the strong ICME category. It should be noted that Medium orbit satellites at $20000$ km are inside the inner magnetosphere if the IMF is Northward during strong ICME space weather conditions. Medium orbit satellites at $10000$ km are protected by the magnetosphere during strong ICME space weather conditions, although exposed to the direct impact of the SW during super-ICME space weather conditions particularly for the Southward IMF orientation.

\subsection{Earth habitability along the Early Sun main sequence}

This section is dedicated to the analysis of the Earth habitability with respect to the space weather conditions along the Sun evolution in the main sequence. Early stages of the Sun evolution are linked to a faster rotation rate and a higher magnetic activity \citep{Emeriau}, because the Sun rotation and magnetic activity decreases during the main sequence \citep{Folsom,Fabbian}. Consequently, the space weather conditions change \citep{Reville,Carolan,Ahuir}. If we consider the rotation tracks from \citep{Carolan} (table 1), the average values of the SW radial velocity, density and radial IMF intensity for different Sun rotation rates are provided, thus the effect of the space weather conditions on the Earth magnetosphere during different stages of the Sun evolution among the main sequence can be studied in the first approximation. Regarding the actual rotation rate of the Sun ($\Omega_{S}$), the rotation rate decreased from $10 \Omega_{S}$ to $2 \Omega_{S}$ between the first $300$ to $1100$ Myr of the Sun evolution. The SW density approximately decreases from $1000$ to $140$ cm$^{-3}$, the radial velocity from $1100$ to $700$ km/s ($P_{d}$ drops from $980$ to $57$ nPa) and $|B|_{IMF}$ from $150$ to $7$ nT. Table \ref{7} shows the averaged magnetopause stand off distance for a Southward IMF during regular space weather condition with respect to $ \Omega_{S}$ based on the parametric study ($R_{sd} / R_{E} = A |B|_{IMF}^{\alpha} P_{d}^{\beta}$). The uncertainty of $R_{sd} / R_{E}$ is calculated as $\Delta (R_{sd} / R_{E}) = |B|_{IMF}^{\alpha} P_{d}^{\beta} \Delta A + \alpha A P_{d}^{\beta} |B|_{IMF}^{\alpha - 1} \Delta \alpha + \beta A |B|_{IMF}^{\alpha} P_{d}^{\beta -1} \Delta \beta$.

\begin{table}[h]
\centering
\begin{tabular}{c | c c c }
IMF & $R_{sd} / R_{E}$ ($2\Omega_{S}$) & $R_{sd} / R_{E}$ ($5\Omega_{S}$) & $R_{sd} / R_{E}$ ($10\Omega_{S}$) \\ \hline
Earth-Sun & $10$ & $4.1$ & $2.2$ \\ 
 & $\pm 2$ & $\pm 0.8$ & $\pm 0.4$\\ 
Northward & $7.1$ & $4.1$ & $2.8$ \\
 & $\pm 0.5$ & $\pm 0.3$ & $\pm 0.2$\\ 
Southward & $5.6$ & $2.5$ & $1.44$ \\
 & $\pm 0.4$ & $\pm 0.2$ & $\pm 0.11$\\ 
Ecliptic & $6.4$ & $3.2$ & $1.94$ \\
 & $\pm 0.6$ & $\pm 0.3$ & $\pm 0.18$\\ 
\end{tabular}
\caption{Averaged magnetopause stand off distance for a Southward IMF during regular space weather condition with respect to $ \Omega_{S}$. Standard error of the regression parameters and derived $R_{sd} / R_{E}$ uncertainty included.}
\label{7}
\end{table}

\begin{figure}[h]
\centering
\resizebox{\hsize}{!}{\includegraphics[width=\columnwidth]{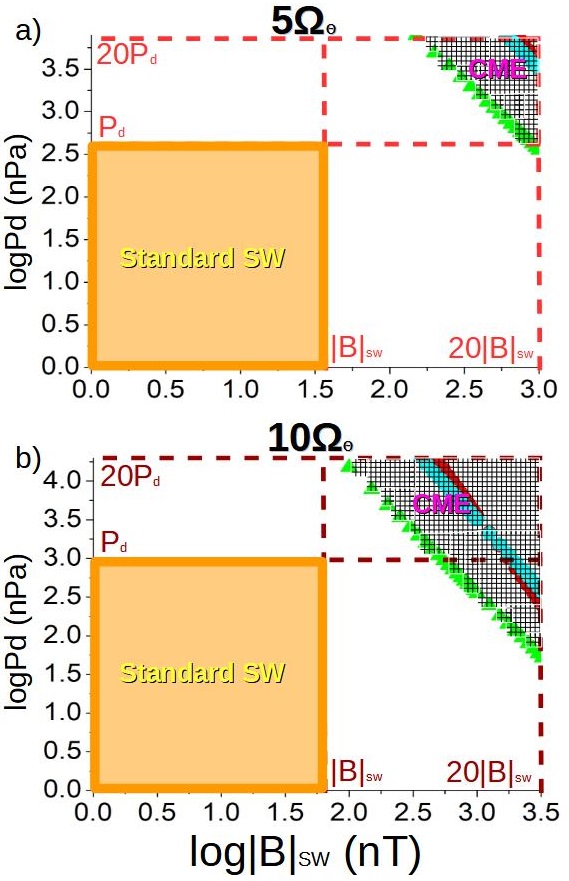}}
\caption{Critical $P_{d}$ and $|B|_{IMF}$ values for the direct precipitation of SW toward the Earth surface for a Southward IMF if (a) $5 \Omega_{S}$ and (b) $10 \Omega_{S}$. The gray dashed region indicates the space weather conditions that lead to the direct precipitation of the SW toward the Earth surface. The orange dashed region shows the regular space weather conditions. Fixed $T = 1.8 \cdot 10^{5}$ K.}
\label{20}
\end{figure}

The analysis results indicate that the Earth magnetosphere avoids the direct precipitation of the SW during regular space weather condition if $\Omega = 5 - 10 \Omega_{S}$, consistent with \citep{Carolan} analysis. Nevertheless, super CMEs during the early main sequence of the Sun should be more frequent, intense and last longer (2 or 3 days) \citep{Sterenborg,Airapetian2,Airapetian3} as the detection of super-flares by Kepler mission for main sequence G-K stars suggests \citep{Shibayama}. Figure \ref{20} shows the range of parameters required for the direct precipitation of the SW during regular (orange dashed surface) and CME (gray dashed surface) space weather conditions in the early main sequence of the Sun ($\Omega = 5 - 10 \Omega_{S}$ and $< 440 Myr$). For simplicity, the space weather conditions consistent with CMEs among the Sun Early main sequence are in the range of $1$ to $20$ times the averaged $P_{d}$ and $|B|_{IMF}$ values provided by \citep{Carolan} (dashed lines). Only the data for Southward IMF orientations is shown, leading to the most restrictive space weather conditions.

The analysis shows a possible direct precipitation of the SW during early phases of the Sun main sequence for CME-like space weather conditions, particularly if the Sun rotation rate is $10 \Omega_{S}$, indicating a wide range of parameters leading to $R_{sd} / R_{E} < 1$.

In summary, the emerging life at the Earth surface is protected from the sterilizing effect of the SW during regular and CME-like space weather conditions $1100$ Myr after the Sun enters in the main sequence. On the other hand, during the first $440$ Myr of the Sun main sequence, CME and super-CME are a major hazard for the Earth habitability, specially due to the high recurrence and prevalence of extreme space weather conditions. It should be noted that the present study results are comparable with the studies by \citep{See2,Airapetian4}.

\subsection{Simulation of ICMEs impacting the Earth between $1997-2020$}

A set of simulations is performed reproducing the space weather conditions during the ICMEs that reached the Earth between the years $1997-2020$. The CMEs included in the analysis are a sub-sample of the CME Richardson list \citep{Richardson2} selecting the most extreme events with respect to $P_{d}$ and $|B|_{IMF}$ values, listed in the Appendix D. Regarding the proposed ICME classification, all the event are Common ICME except the space weather conditions of the $16/07/2000$ and $24/11/2001$ dates, close to the Strong ICME category ($P_{d} \approx 30$ nPa and $|B|_{IMF} \approx 50$ nT). Table \ref{8} indicates the location of the nose of the Bow Shock, magnetopause stand off distance and the lowest OCB latitude at the DS and NS in the North and South Hemispheres for the ICMEs analyzed. In addition, the $K_{p}$ index is calculated from the lowest latitude of OCB line at the North Hemisphere. It should be noted that the IMF and SW values used as the simulation input represent the combination of SW dynamic pressure, IMF module and Southward IMF component that causes the strongest disturbance of the Earth magnetosphere during the ICME, although not necessarily the largest dynamic pressure, IMF module and Southward IMF component because these maxima may occur at different times. Figures \ref{x2} and \ref{x3} in the Appendix D show the 24 hours evolution of the IMF module and components, the velocity module and radial component as well as the SW density and temperature for the $15/05/1997$ and $31/03/2001$, respectively. The model resolution in the latitudinal direction is doubled, reducing the uncertainty of the OCB line from $4^{o}$ to $2^{o}$, thus the uncertainty of the calculated $K_{p}$ index is $\pm 1$.

\begin{table}[h]
\centering
\begin{tabular}{c | c c c c}
Date & $\frac{R_{sd}}{R_{E}} , \frac{R_{BS}}{R_{E}}$ & OCB N-S & OCB N-S & $K_{p}$\\
(dd/mm/yyyy) &   & DS (lat $^{o}$) & NS (lat $^{o}$) \\ \hline
$15/05/1997$ & $6.24-9.97$ & $57-58$ & $58-57$ & $5$ \\
$22/10/1999$ & $3.89-6.75$ & $57-58$ & $56-56$ & $6$ \\ 
$16/07/2000$ & $3.49-5.93$ & $61-61$ & $58-58$ & $5$ \\ 
$31/03/2001$ & $3.53-9.71$ & $63-63$ & $57-58$ & $5$ \\  
$31/03/2001b$ & $4.01-6.34$ & $54-55$ & $55-55$ & $7$ \\   
$24/11/2001$ & $3.27-5.52$ & $54-54$ & $55-55$ & $7$ \\ 
$29/05/2003$ & $3.38-5.52$ & $55-56$ & $56-55$ & $6$ \\ 
$24/10/2003$ & $4.09-6.75$ & $58-58$ & $58-57$ & $5$ \\  
$20/11/2003$ & $4.29-11.66$ & $54-55$ & $56-55$ & $7$ \\  
$07/11/2004$ & $3.48-6.80$ & $60-60$ & $58-59$ & $5$ \\
$21/01/2005$ & $3.75-5.27$ & $55-55$ & $56-56$ & $6$ \\ 
$15/05/2005$ & $3.88-6.59$ & $54-54$ & $55-55$ & $7$ \\ 
$24/08/2005$ & $3.52-6.63$ & $53-53$ & $55-54$ & $7$ \\ 
$24/10/2011$ & $5.82-9.15$ & $56-56$ & $56-56$ & $6$ \\
$13/11/2012$ & $5.10-7.98$ & $57-58$ & $58-58$ & $5$ \\ 
$17/03/2015$ & $5.30-8.49$ & $57-58$ & $58-58$ & $6$ \\  
$03/08/2016$ & $6.91-12.53$ & $58-58$ & $58-58$ & $5$ \\ 
$27/05/2017$ & $5.33-7.74$ & $58-58$ & $58-58$ & $5$ \\ 
$16/07/2017$ & $5.53-9.11$ & $57-58$ & $59-58$ & $5$ \\ 
$20/04/2020$ & $5.94-8.80$ & $59-60$ & $60-58$ & $4$ \\ 
\end{tabular}
\caption{ICME date (first column), magnetopause and Bow Shock stand off distance (second column), OCB minimum in the North and South Hemispheres at the day side (third column) and night side (fourth column) and $K_{p}$ index derived from the lowest OCB line latitude at the North Hemisphere (fifth column).}
\label{8}
\end{table}

\begin{figure*}[h]
\centering
\resizebox{\hsize}{!}{\includegraphics[width=\columnwidth]{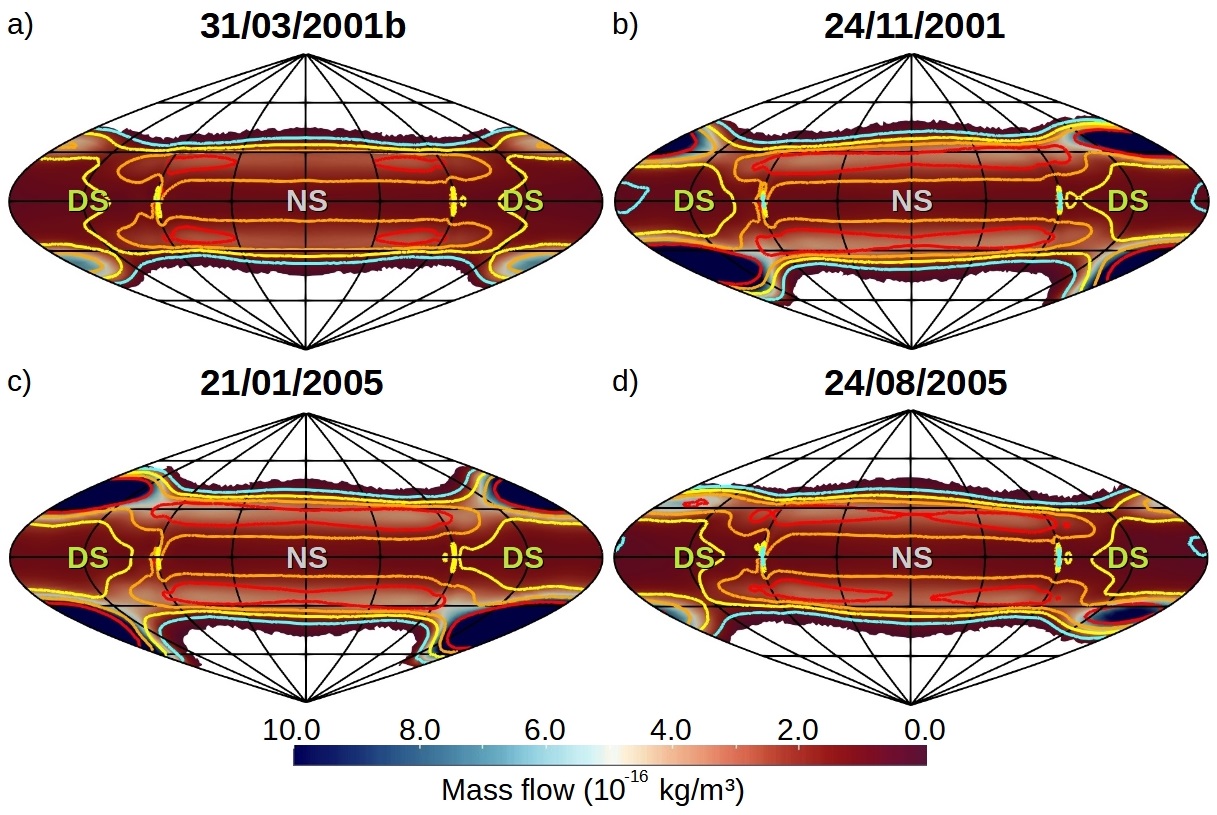}}
\caption{Plasma flow (color surfaces $10^{-16}$ kg/m$^{3}$) at $R/R_{E} = 2.75$ for the ICMEs that impacted the Earth the (a) $31/03/2001b$, (b) $24/11/2001$, (c) $15/05/2005$ and (d) $24/08/2005$. Velocity isocontour of $50$ km/s (light cyan), $100$ km/s (yellow), $150$ km/s (orange) and $200$ km/s (red).}
\label{21}
\end{figure*}

The ICMEs that lead to the smallest magnetopause stand off distance impacted the Earth on $24/11/2001$ and $29/05/2003$ with $R_{sd} / R_{E} < 3.4$. The lowest OCB latitudes at the North and South Hemispheres, $< (53^{o}$, are observed for the ICMEs of the dates $31/03/2001b$, $24/11/2001$, $15/05/2005$ and $24/08/2005$, that is to say, the North of Canada, Alaska, North of Russia and the Nordic countries (except continental Denmark) are exposed to the plasma precipitation along the open magnetic fields. The calculated $K_{p}$ index derived from the lowest OCB line latitude at the North Hemisphere is consistent with the measured $K_{p}$ considering the $\pm 1$ uncertainty, except for the $29/05/2003$ ICME that shows a calculated $K_{p}$ index two units smaller regarding the measured index. It should be noted that the aurora is generated by the electron and ions precipitating towards the Earth surface, thus the plasma flows in the simulations must be also analyzed. Figure \ref{21} indicates the plasma flows and velocity isocontours of the inflow regions at $R/R_{E} = 2.75$ for the ICMEs of the dates $31/03/2001b$, $24/11/2001$, $15/05/2005$ and $24/08/2005$ (ICMEs with the largest observed $K_{p} = 8$ index). In addition, table \ref{9} shows the latitude of the plasma flow extrapolated to the Earth surface at the North and South Hemispheres (inflow plasma velocity $\geq 50 km/s$).

\begin{table}[h]
\centering
\begin{tabular}{c | c c c c}
Date & DS N & DS S & NS N & NS S \\
(dd/mm/yyyy) &  (lat $^{o}$) & (lat $^{o}$) &  (lat $^{o}$) & (lat $^{o}$) \\  \hline
$31/03/2001b$ & $52-63$ & $52-66$ & $55-62$ & $54-61$ \\           
$24/11/2001$ & $51-66$ & $51-72$ & $55-63$ & $55-63$ \\              
$21/01/2005$ & $51-69$ & $51-74$ & $55-62$ & $56-63$ \\              
$24/08/2005$ & $50-62$ & $50-66$ & $54-63$ & $55-62$ \\              
\end{tabular}
\caption{ICME date (first column), plasma flows latitude at the DS North Hemisphere (second column) and South Hemisphere (third column), NS North Hemisphere (fourth column) and South Hemisphere (fifth column) Inflow plasma velocity $\geq 100 km/s$).}
\label{9}
\end{table}

The simulations indicate the formation of plasma streams connecting the Magnetosheath and the Earth surface (see fig \ref{21}, blue color and isolines). Regarding the IMF orientation, the inflow plasma regions in the Earth day side show an East-West asymmetry caused by the large IMF component on the Ecliptic plane, for example for the $24/11/2001$ ICME, or a North-South asymmetry due to the IMF component on the Sun-Earth direction, for example for the $24/08/2005$ ICME. Extrapolating the plasma flows to the Earth surface ($v \ge 50$ km/s, cyan line isoline), the plasma streams are deposited at the planet day side between the latitudes $50-74^{o}$. The simulation of the $24/08/2005$ ICME shows the plasma deposition in the lowest latitude at the North Hemisphere DS $50-62^{o}$, although the simulation of the $21/01/2005$ ICME indicates the widest plasma deposition region at the North Hemisphere DS $51-69^{o}$. Consequently, the plasma streams scatter from the OCB line and deposit at the Earth surface from lower latitudes (between $2-3^{o}$ below the OCB line latitude). If the $K_{p}$ index is calculated including the scattering of the plasma streams, the $K_{p}$ index for the simulations analyzed is $8$, the same value with respect to the measured  $K_{p}$ index. Regarding the plasma flows towards the Earth NS, the inflow maxima ($v \ge 200$ km/s, dark blue isoline) is observed between $54-56^{o}$ latitude, consistent with the latitude of the Aurora observations during extreme space weather conditions \citep{Shaw,Hayakawa,Mikhalev}. It should be noted that the plasma flows for the rest of ICME simulations is smaller with respect to the highlighted cases.

\begin{figure}[h]
\centering
\resizebox{\hsize}{!}{\includegraphics[width=\columnwidth]{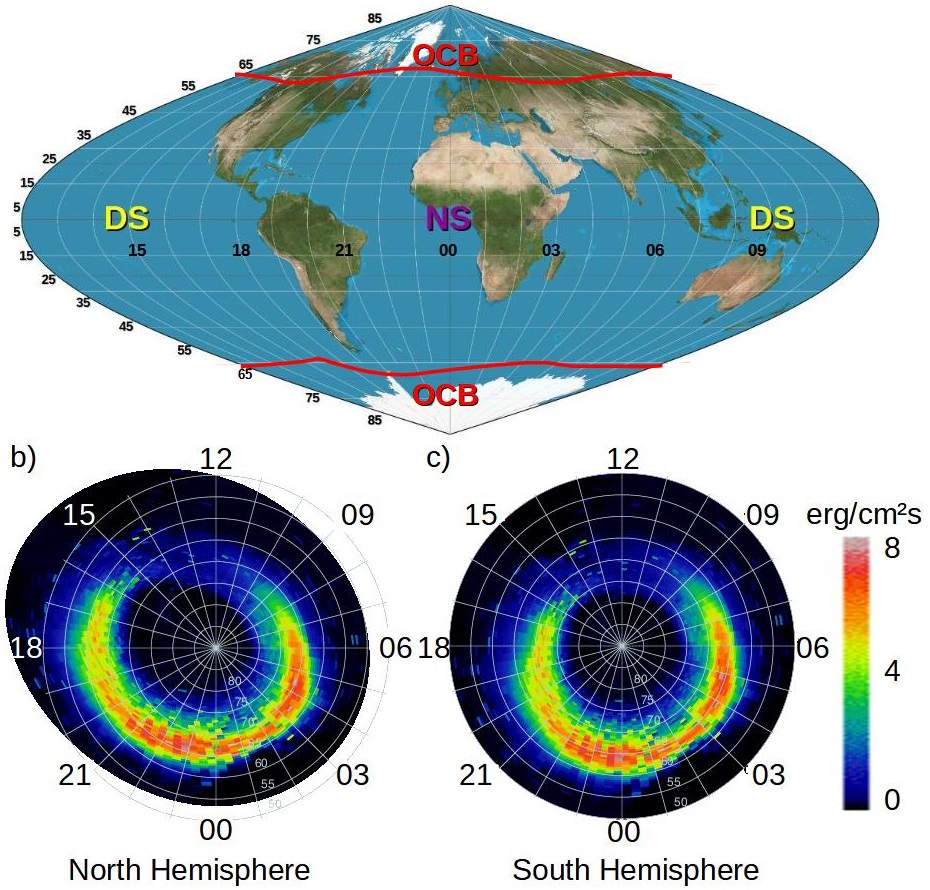}}
\caption{(a) OCB line calculated from the $27/05/2017$ ICME simulation. Energy flux calculated by Ovation prime simulations at the (b) North Hemisphere and (c) South Hemisphere. Ovation prime data is provided by the iNTEGRATED SPACE WEATHER ANALYSIS SYSTEM (iSWA) \citep{iSWA}.}
\label{22}
\end{figure}

Figure \ref{22} shows the OCB line calculated for the $27/05/2017$ ICME simulations and the energy flux calculated by Ovation Prime simulations the date $27/05/2017$ $t = 22:00$ hours at the North and South Hemisphere. The Ovation Prime model is used to forecast the latitude and longitude of the visible aurora \citep{Newell,Machol}. Ovation prime data is provided by the iNTEGRATED SPACE WEATHER ANALYSIS SYSTEM (iSWA) \citep{iSWA}.

The latitude of the OCB line in the Earth night side is located between $58-64^{o}$ at the North Hemisphere and $59-65^{o}$ at the South Hemisphere, similar to the Ovation prime forecast of the aurora location that indicates a local maxima of the energy flux between the latitudes $58-65^{o}$. In addition, the $K_{p}$ index derived from the OCB line is $5$, the same value with respect to the measured $K_{p}$ index the day $27/05/2017$ at $t = 22:00$ hours.

\begin{figure}[h]
\centering
\resizebox{\hsize}{!}{\includegraphics[width=\columnwidth]{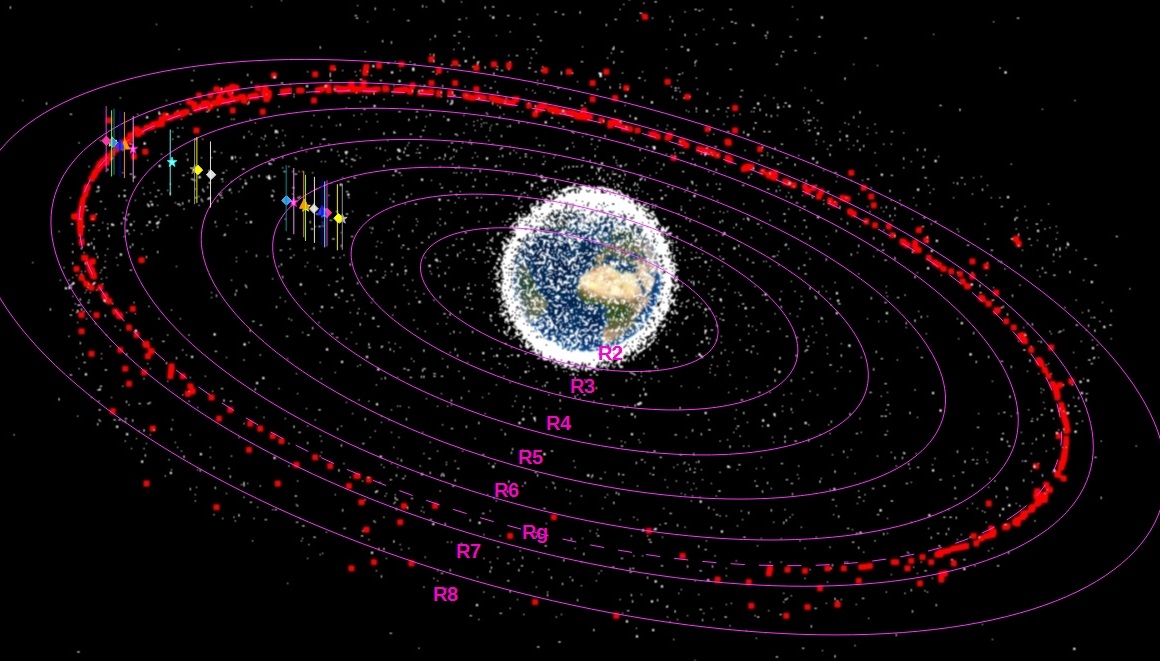}}
\caption{Schematic view of the satellites at Geosynchronous, High, Medium and Low orbits around the Earth between $2000-2020$. The highlighted orbits by red dots are Geosynchronous satellites. The pink solid lines indicate orbits at $R/R_{E} = i$ with $i$ an integer between $2$ to $8$ ($R_{i}$). The pink dashed line indicates the Geosynchronous orbit ($R_{g}$). The colored symbols and horizontal lines show the stand off distance of the magnetopause and the Bow Shock nose for the different ICMEs analyzed.}
\label{23}
\end{figure}

Figure \ref{23} shows the satellites at Geosynchronous, High, Medium and Low orbits around the Earth with respect to the location of the Bow Shock nose and magnetopause stand off distance calculated from the ICME simulation results, particularly for the space weather conditions leading to the lowest magnetopause stand off distance, threatening the satellite integrity by a direct exposition to the SW.

The space weather conditions for the $16/07/2000$, $31/03/2001b$, $24/11/2001$, $29/05/2003$, $21/01/2005$ and $15/05/2005$ ICMEs lead to $R_{BS} < R_{go}$, that is to say, the Geosynchronous satellites face directly the SW during a fraction of the orbit at the Earth day side. Consequently, the integrity of Geosynchronous satellites is endangered specially because of the high SW density, up to $50$ cm$^{-3}$. On the other hand, the space weather conditions during the $03/08/2016$ ICMEs lead to $R_{sd} > R_{go}$ thus the Geosynchronous satellites are inside the Earth inner magnetosphere during the full orbit. The space weather conditions during the rest of ICME analyzed show a $R_{BS} > R_{go} > R_{sd}$, thus the Geosynchronous satellites are inside the Earth magnetosphere along the full orbit, although for a fraction of the orbit at the day side, the satellites cross the Magnetosheath and enter inside the Bow Shock. Inside the Bow Shock the SW particles slow down and accumulate, leading to a decrease of the protection the magnetosphere brings. The model developed by \citet{Dmitriev} predicts the magnetopause crossing of geosynchronous satellite during the $20/11/2003$, $07/11/2004$, $15/05/2005$ and $24/08/2005$ ICMEs, consistent with the simulation results that indicate a magnetopause stand off distance below the geosynchronous satellite orbit: $4.29$, $3.48$, $3.88$ and $3.52$ $R_{E}$, respectively. In addition, there are observations of magnetopause crossing by the geosynchronous satellite $1991-080$, $1994-084$ and $LANL-01A$ during $31/03/2001$ ICME \citep{Ober}, by GOES $13 - 15$ and MMS during $17/03/2001$ ICME \citep{Le}, by THEMIS A and E during $27/05/2017$ ICME \citep{Pezzopane}, also consistent with the simulation results that predict a magnetopause stand off distance of $4.01$, $5.30$ and $5.33$ $R_{E}$, respectively. It should be noted that the satellites at Medium orbits below $10000$ km are inside the magnetosphere along the full orbit for all the ICMEs analyzed, although Medium orbits at $20000$ km cross the Magnetosheath during the $22/10/1999$, $31/03/2001$, $24/10/2003$, $07/11/2004$ and $24/08/2005$ ICMEs. An example of the consequences of the severe space weather conditions during the $24/10/2003$ ICME were the loss of the Low orbit satellite ADEOS/MIDORI 2 due to electrostatic discharge, the engine switch off the high orbit satellite SMART-1 caused by the ionization effect of the solar wind, two weeks outage of the geostationary satellite DRTS/Kodama also caused by electrostatic discharge as well as high bit error rates and magnetic torques disabled of the GOES $9$, $10$ and $12$ \citep{Tamaoki,Cannon}. The simulation of the $24/10/2003$ ICME predicts a minor-moderate auroral activity, a significant East-West tilt of the Earth magnetosphere as well as a relatively low magnetopause stand off distance ($R_{sd}/R_{E} = 4.09$).

\section{Discussion and conclusions}
\label{Conclusions}

The distortion induced by the solar wind and the interplanetary magnetic field on the Earth magnetosphere topology must be analyzed in detail, because the large variability of the space weather conditions leads to the triggering of a wide number of physical phenomena, for example geomagnetic storms. Extreme space weather conditions have consequences in the integrity of satellites orbiting the Earth, electric power grids and humans health. In addition, an efficient shielding of the exoplanets magnetic field to avoid the direct precipitation of the stellar wind toward the surface is critical for the exoplanet habitability.   

Simulate the interaction of the SW and IMF with the Earth magnetic field using MHD models is an useful tool to analyze the global structures of the magnetosphere during different space weather conditions. It should be noted that the validity of the MHD models was confirmed comparing simulation results and spacecraft / ground based measurements \citep{Watanabe,Raeder2,Wang6,Facsko,Honkonen}. Consequently, a parametric analysis based on MHD simulations regarding the SW density, velocity and temperature as well as the IMF orientation and intensity may provide a reasonable first approximation of the space weather effects on the Earth magnetosphere topology. We recall no kinetic effects are included in the study, thus deviations between simulation results and observational data can exist for some of the extreme space weather configurations analyzed.

The set of simulations performed fixing the SW dynamic pressure although modifying the IMF orientation and intensity show the critical role of the IMF on the Earth magnetosphere topology, leading to a large variation on the magnetopause stand off distance, the location of the reconnection regions between the IMF and the Earth magnetic field where the SW is injected inside the inner magnetosphere, the plasma streams between the Magnetosheath and the Earth surface as well as the open-close field line boundary. Particularly, the Southward, Sun-Earth and Earth-Sun IMF orientations lead to the smallest magnetopause stand off distances as the IMF intensity increases. In addition, the reconnection regions are closer to the Earth surface as the IMF intensity increases, although at different locations inside the magnetosphere regarding the IMF orientation, modifying the plasma flows towards the Earth surface. The same way, an intense IMF oriented in the Southward direction causes a decrease of the latitude of the open-close field line boundary, exposing wider regions of the Earth surface to the plasma flows along the magnetic field lines. For example, the open-close field line boundary at the day side ($0^{o}$ longitude) decreases from $72$ to $53^{o}$ if the simulations with no IMF and Southward IMF with $250$ nT are compared. It is worth stressing that the results of the parametric study is consistent with other authors results regarding the magnetopause stand off distance \citep{Song,Kabin,Lavraud,Ridley,Meng,Wang5} and the latitude of the open-close field line boundary \citep{Lopez,Kabin,Wild,Rae,Wang7,Burrell}.

The simulations indicate the direct precipitation of the SW toward the Earth surface is very unlikely in the range of space weather conditions expected for the present and future stages of the Sun along the main sequence evolution. The extreme space weather conditions during ICME and super-ICME impacting the Earth cannot lead to a compression and erosion of the Earth magnetosphere large enough to reduce the magnetosphere stand off distance below $R_{sd}/R_{E} = 1$. For example, the SW precipitation towards the Earth surface for a IMF purely oriented in the Southward direction requires a IMF intensity of $1000$ nT and a SW dynamic pressure above $350$ nPa, space weather conditions well above super-ICMEs. In addition, if the analysis is extended to previous stages of the solar evolution, the simulations show an efficient shielding of the Earth surface $1100$ Myr after the Sun enters in the main sequence. On the other hand, early evolution stages once the Sun rotation rate was $5-10$ times faster with respect to the present rotation rate, the case of the first $440$ Myr of the Sun main sequence, the Earth habitability could be threatened during extreme space weather conditions, results consistent with \citep{See2,Airapetian4} studies. It should be noted that the Earth magnetic field intensity is a fixed parameter in the analysis, although several studies indicates the Earth magnetic field could be stronger during early evolution phases of the solar system \citep{Tarduno,Tarduno2,Tarduno3}. In addition, there are other factors that affect the young Earth habitability not included in the study, for example the Sun luminosity and X ray / ultra violet emission \citep{Cockell,Sackmann,Ribas,Cnossen} or the atmosphere evolution \citep{Kasting2,Arndt,Gronoff}. Consequently, supplementary analysis are required to confirm the habitability of the young Earth, that will be the topic for a future research.

An ICME classification for the Earth is proposed regarding the SW dynamic pressure, IMF intensity and Disturbance Storm Time Index. The ICME classification consist of three categories: Common, Strong and Super ICMEs. Common ICMEs have a relatively large recurrence covering the main number of extreme space weather events impacting the Earth with SW dynamic pressures $< 40$ nPa, IMF intensities $< 50$ nT and $Dst < -100$ nT. Strong ICMEs have a smaller recurrence, a few events each year, particularly associated with the maximum of the Sun magnetic activity cycle, showing a dynamic pressure in the range of $[40,100]$ nPa, IMF intensity of $[50,100]$ nT and $Dst = [-100,-500]$ nT. Super-ICME category identifies once per century events similar to the 'Carrington' event with a SW dynamic pressure above $100$ nPa, IMF intensity above $100$ nT and $Dst > -500$ nT. Present classification is consistent with other author studies as \citep{Rastatter,Tsurutani,Siscoe,Saiz,Balan,Keika}. 

The simulations performed to reproduce the effect of the ICMEs impacting the Earth between $1997-2020$ indicate that all the events can be included inside the Common ICME category, except the extreme space weather conditions observed the dates $16/07/2000$ and $24/11/2001$, classified close to the Strong ICME category ($P_{d} \approx 30$ nPa and $|B|_{IMF} \approx 50$ nT). It should be mentioned that $29/10/2003$ event is not included in the study due to the lack of SW and IMF data, although this event should be probably inside the Strong ICME category \citep{Balan}. For example, the Earth magnetosphere distortion during the $24/11/2001$ ICME was large enough to potentially impact the electric grids of North of Canada, Alaska, North of Russia and the Nordic countries (except continental Denmark). In addition, the $K_{p} = 8$ index calculated regarding the plasma flows towards the Earth surface is the same than the $K_{p} = 8$ index measured.

It must be recalled that the simulations neglect the effect imprinted in the Earth magnetosphere by previous space weather conditions. Consequently, the simulations performed for ICMEs showing a fast variation of the space weather conditions could overestimate the forcing of the SW and IMF. This is the case of the simulations performed for the $31/03/2001$ ICME, see figure \ref{x3} in the Appendix D, showing large variations of the IMF orientation and intensity as well as SW dynamic pressure in the time frame of minutes. On the other hand, the space weather parameters are quasi-steady in the time frame of $1$ hour during the $15/05/1997$ ICME, see figure \ref{x2} in the Appendix D, thus the simulation results are more accurate.

Despite the fact that no direct SW precipitation is expected toward the Earth surface, extreme space weather conditions can endanger the integrity of the satellites around the Earth, because the magnetopause stand off distance decreases and the satellite orbit at the Earth day side is partially unprotected outside the magnetosphere. Southward and Ecliptic IMF orientations are particularly adverse for Geosynchronous satellites, partially exposed to the SW if the SW dynamic pressure is $\approx 14-26$ nPa and the IMF intensity $10$ nT, that is to say, $5-10$ times the dynamic pressure of regular space weather conditions. On the other hand, Medium orbit satellites at $20000$ km are directly exposed to the SW during Common ICME if the IMF orientation is Southward and during Strong ICME if the IMF orientation is Earth-Sun or Ecliptic. The same way, Medium orbit satellites at $10000$ km are directly exposed to the SW if a Super ICME with Southward IMF orientation impacts the Earth. For example, during the ICMEs of the dates $15/07/2000$, $24/11/2001$, $29/05/2003$ and $21/01/2005$ the Geosynchronous satellites suffered the direct impact of the SW during a fraction of the orbit at the Earth day side, although Medium orbit satellites below $R/R_{E} \approx 5$ were protected by the magnetosphere among the full orbit. It should be noted that other important threats to the satellite integrity during extreme space weather conditions, as the enhancement of the Earth radiation belts and the atmosphere drag force, are not included in the study.

\begin{acknowledgements}
This work was supported by the project 2019-T1/AMB-13648 founded by the Comunidad de Madrid. The research leading to these results has received funding from the grants ERC WholeSun, Exoplanets A and PNP. We extend our thanks to CNES for Solar Orbiter, PLATO and Meteo Space science support and to INSU/PNST for their financial support. This work has been supported by Comunidad de Madrid (Spain) - multiannual agreement with UC3M (“Excelencia para el Profesorado Universitario” - EPUC3M14 ) - Fifth regional research plan 2016-2020. Data available on request from the authors. The authors acknowledge the Community Coordinated Modeling Center (CCMC) and the main developer of Ovation prime code Prof. Patrick Newell.
\end{acknowledgements}

\begin{appendix}

\section{Upper ionospheric model}

The upper ionospheric domain is located between $R = 2 - 2.5 R_{E}$ in simulations with $P_{d} > 1$ nPa and between $R = 3 - 3.5 R_{E}$ in simulations with $P_{d} < 1$ nPa. The upper ionospheric model is based on \citep{Buchner}. Below the lower boundary of the upper ionosphere the magnetic field intensity is too large thus the simulation time step is too small. In addition, a single fluid MHD model cannot reproduce correctly magnetosphere regions as the inner ionosphere or the plasma-sphere because the kinetic effects are large.

First, the field aligned current ($J_{FAC}$) are calculated as:
\begin{equation}
\label{eqn:1}
\vec{J_{FAC}} = \vec{J} - \vec{J_{\perp}}
\end{equation}
where:
\begin{equation}
\label{eqn:2}
\vec{J} = \frac{1}{mu_{0}} \vec{\nabla} \times \vec{B}
\end{equation}
\begin{equation}
\label{eqn:3}
\vec{J_{\perp}} = \vec{J} - \frac{J_{r} B_{r} + J_{\theta} B_{\theta} + J_{\phi} B_{\phi}}{|B|^{2}} \vec{B}
\end{equation}
with $\vec{J}$ the plasma current, $\vec{J_{\perp}}$ the perpendicular component of the plasma current along the magnetic field line, $mu_{0}$ the vacuum magnetic permeability and $\vec{B}$ the magnetic field.

Next, the electric field of the upper ionosphere model is calculated using the Pedersen conductance ($\sigma$) empirical formula:
\begin{equation}
\label{eqn:4}
\sigma = \frac{40 E_{0} \sqrt{F_{E}}}{16 + E_{0}^{2}}
\end{equation}
with $E_{0} = K_{B} T_{e}$ the mean energy of the electrons, $F_{E} = n_{e} \sqrt{E_{0}/(2 \pi m_{e})}$ the energy flux and $K_{B}$ the Boltzmann constant ($T_{e}$ and $m_{e}$ the electron temperature and mass, respectively). Thus, the electric field ($E$) linked to the FAC is: 
\begin{equation}
\label{eqn:5}
\vec{E} = \sigma \vec{J_{FAC}}
\end{equation}
Once the electric field is calculated, the velocity of the plasma in the upper ionosphere is:
\begin{equation}
\label{eqn:6}
\vec{v} = \frac{\vec{E} \times \vec{B}}{|B|^{2}}
\end{equation}
The plasma density in the upper ionosphere is defined with respect to the Alfv\'{e}n velocity. The module of the Alfv\'en velocity is fixed ($\mathrm{v}_{A} = 8 \cdot 10^{3}$ km/s) to control the simulation time step, thus the density profile between $R = 2 - 2.5 R_{E}$ does not evolve along the simulation, defined as:
\begin{equation}
\label{eqn:7}
\rho = \frac{|B|^{2}}{\mu_{0}v_{A}^{2}}
\end{equation}
The plasma pressure in the upper ionosphere model is defined with respect to the sound speed of the SW ($c_{sw}$) and at the inner boundary ($c_{p}$):
\begin{equation}
\label{eqn:8}
p = \frac{n}{\gamma} \left( \frac{(c_{p} - c_{sw})(r^{3}-R_{s}^{3})}{R_{un}^{3}-R_{s}^{3}} + c_{sw} \right)^{2}
\end{equation}
with $\gamma = 5/3$ the polytropic index, $c_{p} = \sqrt{\gamma K_{B} T_{p}/m_{p}}$ with $T_{p}$ is the plasma temperature at the inner boundary and $c_{sw} = \sqrt{\gamma K_{B} T_{sw}/m_{p}}$ with $T_{sw}$ the SW temperature. Figure \ref{z} shows the profiles of the density and pressure inside the upper ionosphere model for the simulation with $T_{sw} = 1.8 \cdot 10^{5}$ K, $n = 20$ cm$^{-3}$, $|v|=350$ km/s and $|B|_{IMF}=0$.

\begin{figure}[h]
\centering
\resizebox{\hsize}{!}{\includegraphics[width=\columnwidth]{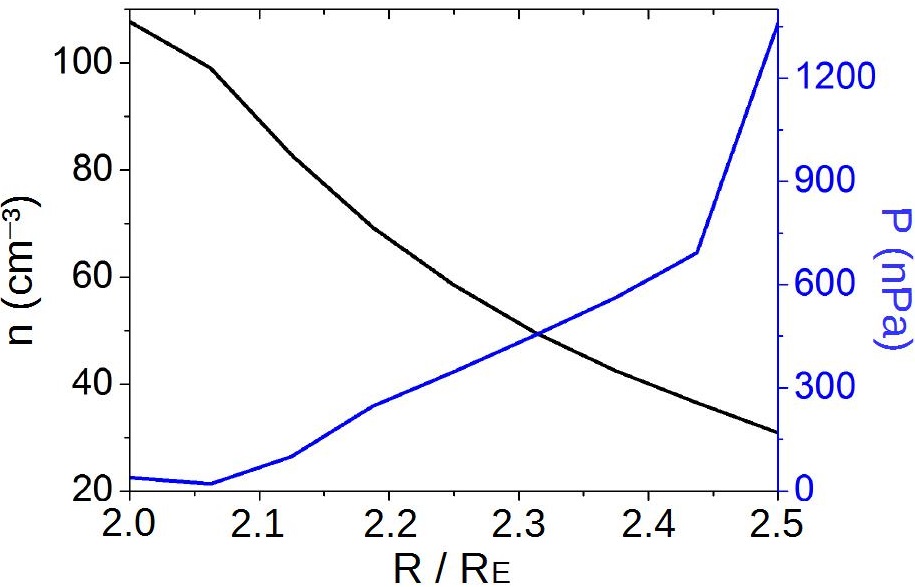}}
\caption{Radial profiles of the density and pressure inside the upper ionosphere model. Simulation with $T_{sw} = 1.8 \cdot 10^{5}$ K, $n = 20$ cm$^{-3}$, $|v|=350$ km/s and $|B|_{IMF}=0$.}
\label{z}
\end{figure}  

The model initial conditions for the plasma density and pressure are defined to have a smooth transition between the upper ionosphere and the simulation domains. Along the simulation the pressure and density gradients increase because the density and pressure profiles are fixed inside the inner ionosphere, although evolving freely in the simulation domain. The answer of the system during the early stages of the simulation is to compensate the gradients feeding plasma towards the simulation domain, generating an outward plasma flux that saturates once the inner magnetosphere reaches the steady state. Henceforth, the plasma flows are dominated by the balance between the solar wind injection inside the inner magnetosphere and the plasma streams towards the planet surface.

The numerical model used to study space weather configurations with low SW dynamic pressure, $P_{d} \le 1$ nPa, is modified with respect to high $P_{d}$ simulations. The inner boundary is located at $R_{in} = 3R_{E}$ and the upper ionosphere domain between $3.0 - 3.5R_{E}$. The reason of this modification is avoiding an overestimation of the magnetosphere thermal pressure in low $P_{d}$ simulations, caused by the plasma Alfv\'en velocity imposed in the upper ionosphere, required to control the simulation time step, leading to an artificial enhancement of the plasma fluxes toward the simulation domain. This numerical issue is avoided displacing outward the inner boundary of the model, reducing the fluxes and minimizing the overestimation of the magnetosphere thermal pressure. It should be noted that, for high $P_{d}$ simulations, the effect of outward fluxes is negligible in the pressure balance.

\begin{figure}[h]
\centering
\resizebox{\hsize}{!}{\includegraphics[width=\columnwidth]{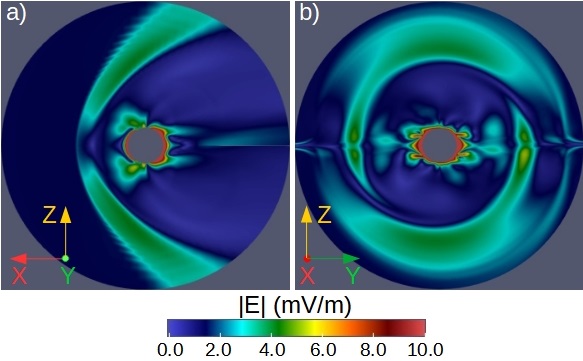}}
\caption{Electric field module in the simulation domain for the benchmarking case in \citet{Samsonov3} at (a) the XZ plane and (b) YZ plane.}
\label{a}
\end{figure}

The model prediction for quite space weather conditions is compared with \citet{Samsonov3} analysis, performing a simulation using the same parameters with respect to the original benchmarking study: $n = 5$ cm$^{-3}$, $V_{x} = -400$ km/s, $T = 2 \cdot 10^{5}$ K, $B_{y} = -B_{x} = 3.5$ nT and $B_{z} = 0$ nT. The location of the magnetopause is: $R_{x}/R_{E} = 10.7$, $R_{y}/R_{E} = 16.8$, $R_{-y}/R_{E} = 16.6$ and $R_{z}/R_{E} = 14.9$. There is a reasonable agreement between the model prediction and the benchmarking study. Figure \ref{a} shows the electric field in the simulation domain. The local maxima of the electric field is also consistent with the simulations in \citet{Samsonov3} near the bow shock (fig 1). It should be noted that there is a secondary local maxima of the electric field module near the lower boundary of the simulation domain caused by the conditions imposed at the upper ionosphere. The module of the electric field predicted inside the magnetosphere is similar to Cluster spacecraft observations during the magnetopause crossing the date $30/02/2002$ \citep{Keyser}. The electric field measured in the current sheet and magnetosheath is one order of magnitude larger with respect to the simulations because the IMF module is $10$ times larger during Cluster magnetopause crossing. If the simulation is performed using a Southward IMF with $|B|=50$ nT and $P_{d} = 5$ nPa, similar to the space weather conditions during Cluster  magnetopause crossing, the electric field predicted is $15-30$ mV/m at the current sheet and magnetosheath region, similar to Cluster spacecraft observations.

In addition, another two simulations are performed using the same SW parameters although for Northward and southward IMF orientations with $|B_{z}|=3$ nT, identifying the displacement of the magnetopause location defined as $\Delta R/R_{E} = Northward(R)/R_{E} - Southward(R)/R_{E}$: $\Delta R_{x}/R_{E} = 0.2$, $\Delta R_{y}/R_{E} = 0.1$ and $\Delta R_{z}/R_{E} = -1.0$. Again, there is a reasonable agreement.

Next, the model is compared with the Carrington-like event analyzed by \citet{Ridley2} that identified a magnetopause stand off distance of $R/R_{E} = 2$ (equal to the lower boundary of the simulation domain) for the parameters $n = 750$ cm$^{-3}$, $V_{x} = -1600$ km/s ($P_{d}=1600$ nPa), $T = 3.5 \cdot 10^{7}$ K, $B_{x} = 150$ nT, $B_{y} = 170$ nT and $B_{z} = 200$ nT. The present model cannot be used to simulate space weather conditions leading to a magnetopause stand off distance below $R/R_{E} = 2.5$, although the extrapolation of the model results predicts $R/R_{E} \approx 1.22$ if $P_{d}=1600$ nPa and $B_{z} = 200$ nT (pure Southward IMF orientation).

\begin{figure}[h]
\centering
\resizebox{\hsize}{!}{\includegraphics[width=\columnwidth]{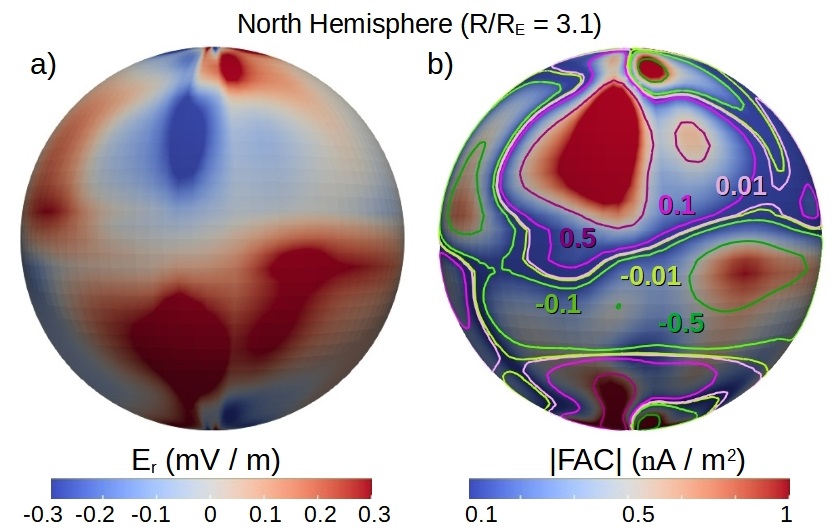}}
\caption{(a) Radial electric field and (b) FAC intensity in the North Hemisphere at $R/RE = 3.1$ for \citet{Samsonov3} benchmarking case. The colored iso-contours indicate different FAC intensities and inward/outward (green/pink) orientations.}
\label{b}
\end{figure}

The electric field in the upper ionosphere domain remains almost unchanged during the simulation because the density profile is fixed. Figure \ref{b} panel a shows the radial electric field inside the upper ionosphere (North hemisphere at $R/RE = 3.1$) for \citet{Samsonov3} benchmarking case, indicating a reasonable order of magnitude agreement with respect to other models and satellite measurements \citep{Shume,Alken,Watanabe2}. Panel b indicates the FAC intensity and orientation, values in the range of the observations and modeling data (from $nA / m^2$ to several $\mu A m^2$ regarding the space weather conditions) \citet{Weimer,Waters,Ritter,Bunescu,Zhang3}.

\section{Magnetic field line mapping}

\begin{figure}[h]
\centering
\includegraphics[width=6cm]{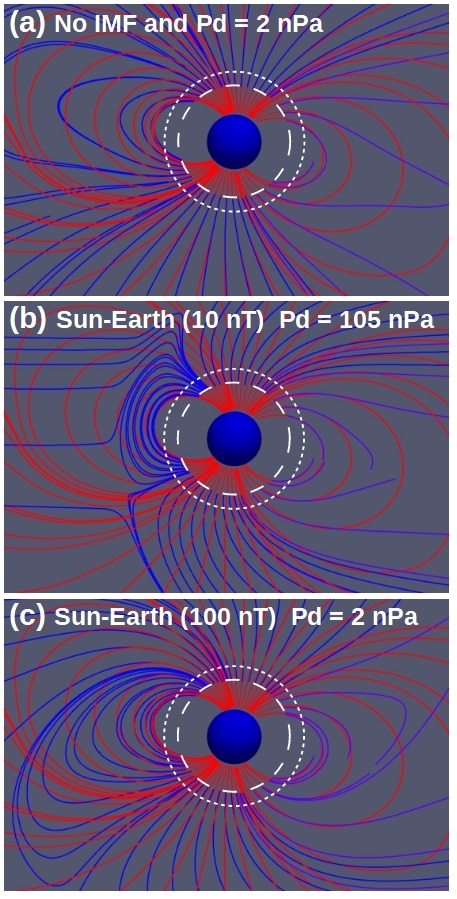}
\caption{3d view of the magnetic field lines mapping between a non-perturbed dipole and simulations with (a) No IMF and $P_{d} = 2$ nPa, (b) Sun-Earth IMF $|B|_{IMF} = 10$ nT and $P_{d} = 105$ nPa and (c) Sun-Earth IMF $|B|_{IMF} = 100$ nT and $P_{d} = 2$ nPa. The dashed white line indicates the inner boundary of the simulation domain ($R / R_{E} = 2$) and the dotted white line the upper boundary of the ionosphere model ($R / R_{E} = 2.5$).}
\label{x}
\end{figure}

The magnetic field lines of the simulations are mapped with respect to the magnetic field of a non perturbed dipole. To that end, a simulation without the driving effect of the SW and IMF is performed decreasing the inner boundary of the simulation domain to $R / R_{E} = 1$. Figure \ref{x} shows the mapping of the magnetic field lines for simulations during regular and extreme space weather conditions with the magnetic field of a non-perturbed dipole.

The mapping shows that, even for the simulations with a large dynamic pressure (panel b) and $|B|_{IMF}$ (panel c), the magnetic field lines follows the non perturbed dipole magnetic field lines inside the computational domain of the upper ionosphere, between $R / R_{E} = 2$ to $2.5$, indicated in the figures by a dashed and a dotted white line, respectively. Consequently, the extrapolations of the OCB line and plasma flows towards the Earth surface are reasonably accurate.

\section{Magnetotail and OCB line latitude at the night side}

The simulation outer boundary is located at $R/R_{E} = 30$, although the magnetotail extension can exceed $R/R_{E} = 30$, thus this magnetosphere structure is only partially reproduced by the model. Consequently, the last close magnetic field line cannot be accurately identified at the Earth night side as well as the latitude of the OCB line. This is the case for the simulations with $P_{d} \ge 85$ nPa and $|B_{IMF}| \le 10$ nT, reason why the analysis of the OCB line latitude at the night side is not performed for such configurations. Nevertheless, the outer boundary conditions could affect the magnetotail topology once the simulation reaches the steady state. Figure \ref{z} compares the magnetotail structure in simulations for the same space weather conditions although increasing the outer boundaries from $R/R_{E} = 30$ to $100$.

\begin{figure}[h]
\centering
\resizebox{\hsize}{!}{\includegraphics[width=\columnwidth]{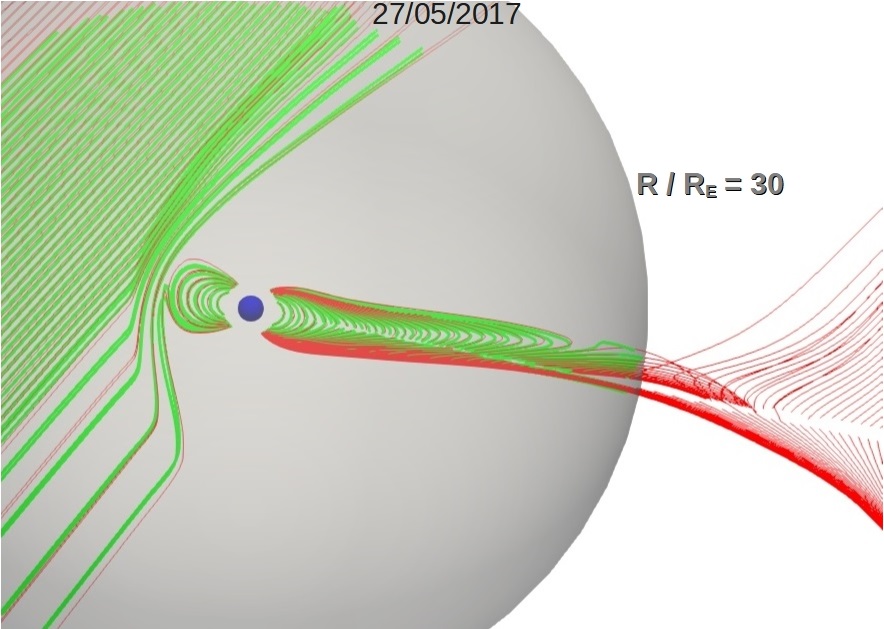}}
\caption{Polar plot of the magnetic field lines in simulations with the outer boundary at $R/R_{E} = 30$ (red line) and $100$ (green line) for the  $27/05/2017$ ICME. The gray sphere indicates the outer boundary of the simulation with at $R/R_{E} = 30$.}
\label{z}
\end{figure}

There is a good agreement between the simulations with the outer boundary at $R/R_{E} = 30$ and $100$ regarding the Earth magnetic field at the day and night side, indicating that the outer boundary conditions have a negligible effect on the computational domain. Consequently, the location of the last close magnetic field lines at the night side is similar, leading to values of the OCB line latitude inside the uncertainty of the model resolution.

\section{CME list}

Table \ref{y} shows the SW density, temperature, radial velocity, dynamic pressure, IMF components and module as well as the data source for the CME sub-sample analyzed.

\begin{table*}[h]
\centering
\begin{tabular}{c | c c c c c c  c c c c}
Date & $n$ & $T$ & $V_{r}$ & $P_{d}$ & $B_{x}$ & $B_{y}$ & $B_{z}$ & $|B|_{sw}$ & Source & $K_{p}$ \\
(dd/mm/yyyy) (hh:hh) & (cm$^{-3}$) & ($10^{3}$ K) & (km/s) & (nPa) & (nT) & (nT) & (nT) & (nT) & \\  \hline
$15/05/1997$ ($07:30$ h) & $25$ & $20$ & $-400$ & $3.34$ & $10$ & $-15$ & $-20$ & $27$ & OMNIWeb & $6$ \\
$22/10/1999$ ($06:30$ h) & $50$ & $40$ & $-550$ & $12.65$ & $0$ & $20$ & $-25$ & $32$ & OMNIWeb & $6$ \\
$16/07/2000$ ($02:20$ h) & $40$ & $50$ & $-1000$ & $33.45$ & $25$ & $30$ & $30$ & $49$ & ACE SWICS + OMNIWeb & $7$ \\
$31/03/2001$ ($01:20$ h) & $25$ & $60$ & $-650$ & $12.37$ & $-50$ & $30$ & $40$ & $71$ & OMNIWeb & $6$ \\
$31/03/2001b$ ($04:20$ h) & $35$ & $30$ & $-700$ & $14.34$ & $0$ & $0$ & $-40$ & $40$ & OMNIWeb & $8$ \\
$24/11/2001$ ($07:10$ h) & $50$ & $40$ & $-850$ & $30.21$ & $-20$ & $-30$ & $-40$ & $54$ & OMNIWeb & $8$ \\
$29/05/2003$ ($19:50$ h) & $50$ & $60$ & $-800$ & $26.76$ & $10$ & $15$ & $-25$ & $31$ & OMNIWeb & $8$ \\
$24/10/2003$ ($18:10$ h) & $50$ & $50$ & $-550$ & $12.65$ & $-20$ & $-20$ & $-15$ & $32$ & OMNIWeb & $5$ \\
$20/11/2003$ ($16:20$ h) & $15$ & $30$ & $-600$ & $4.52$ & $5$ & $30$ & $-45$ & $54$ & OMNIWeb & $8$ \\
$07/11/2004$ ($19:50$ h) & $60$ & $80$ & $-650$ & $21.20$ & $15$ & $-40$ & $30$ & $52$ & OMNIWeb & $6$ \\
$21/01/2005$ ($18:50$ h) & $50$ & $60$ & $-950$ & $37.74$ & $-20$ & $20$ & $-25$ & $38$ & OMNIWeb & $7$ \\
$15/05/2005$ ($06:10$ h) & $25$ & $100$ & $-900$ & $16.93$ & $-30$ & $25$ & $-45$ & $60$ & OMNIWeb & $8$ \\
$24/08/2005$ ($10:10$ h) & $40$ & $40$ & $-750$ & $18.82$ & $-10$ & $35$ & $-55$ & $66$ & ACE SWICS + OMNIWeb & $8$ \\
$24/10/2011$ ($21:00$ h) & $20$ & $20$ & $-500$ & $4.18$ & $10$ & $15$ & $-20$ & $27$ & ACE SWICS + OMNIWeb & $6$ \\
$13/11/2012$ ($00:50$ h) & $40$ & $25$ & $-450$ & $6.77$ & $10$ & $-20$ & $-20$ & $30$ & ACE SWICS + OMNIWeb & $6$ \\
$17/03/2015$ ($06:00$ h) & $23$ & $55$ & $-550$ & $5.82$ & $10$ & $-21$ & $-21$ & $31$ & ACE SWICS + OMNIWeb & $5$ \\
$03/08/2016$ ($05:00$ h) & $15$ & $400$ & $-425$ & $2.27$ & $-2$ & $22$ & $-20$ & $30$ & DSCOVR & $4$ \\
$27/05/2017$ ($21:50$ h) & $60$ & $100$ & $-380$ & $7.25$ & $-10$ & $11$ & $-20$ & $25$ & DSCOVR & $5$ \\
$16/07/2017$ ($09:30$ h) & $30$ & $200$ & $-450$ & $5.08$ & $8$ & $-23$ & $-23$ & $33$ & DSCOVR & $5$ \\
$20/04/2020$ ($08:50$ h) & $35$ & $70$ & $-400$ & $4.68$ & $4$ & $-14$ & $-15$ & $21$ & DSCOVR & $3$ \\
\end{tabular}
\caption{SW and IMF parameters of the CME selection between $1997-2020$. Date (first column), SW density (second column), SW temperature (third column), SW radial velocity (fourth column), SW dynamic pressure (fifth column), IMF component along the Sun-Earth direction (sixth column), IMF component along the magnetic axis direction (seventh column), IMF component along the Ecliptic direction (eighth column), IMF module (ninth column), data source (tenth column) and measured $K_{p}$ index (eleventh column).}
\label{y}
\end{table*}

The simulation inputs are obtained from OMNIWeb \citep{OMNIWeb}, ACE SWICS \citep{ACE} and DSCOVR \citep{DSCOVR} after the front of the ICME impacts the Earth. OMNIWeb provides high resolution OMNI (HRO) data based on the Global Geospace Science (GGS) Wind satellite \citep{Ogilvie}, ACE SWICS data from the Advanced Composition Explorer (ACE) spacecraft \citep{Stone} and DSCOVR data from the Deep Space Climate Observatory \citep{Burt}.

It should be noted that the strong CME impacting the Earth the date $29/10/2003$ is not included in the list because there is not available data of the SW density and temperature neither the IMF module and intensity.

\begin{figure}[h]
\centering
\resizebox{\hsize}{!}{\includegraphics[width=\columnwidth]{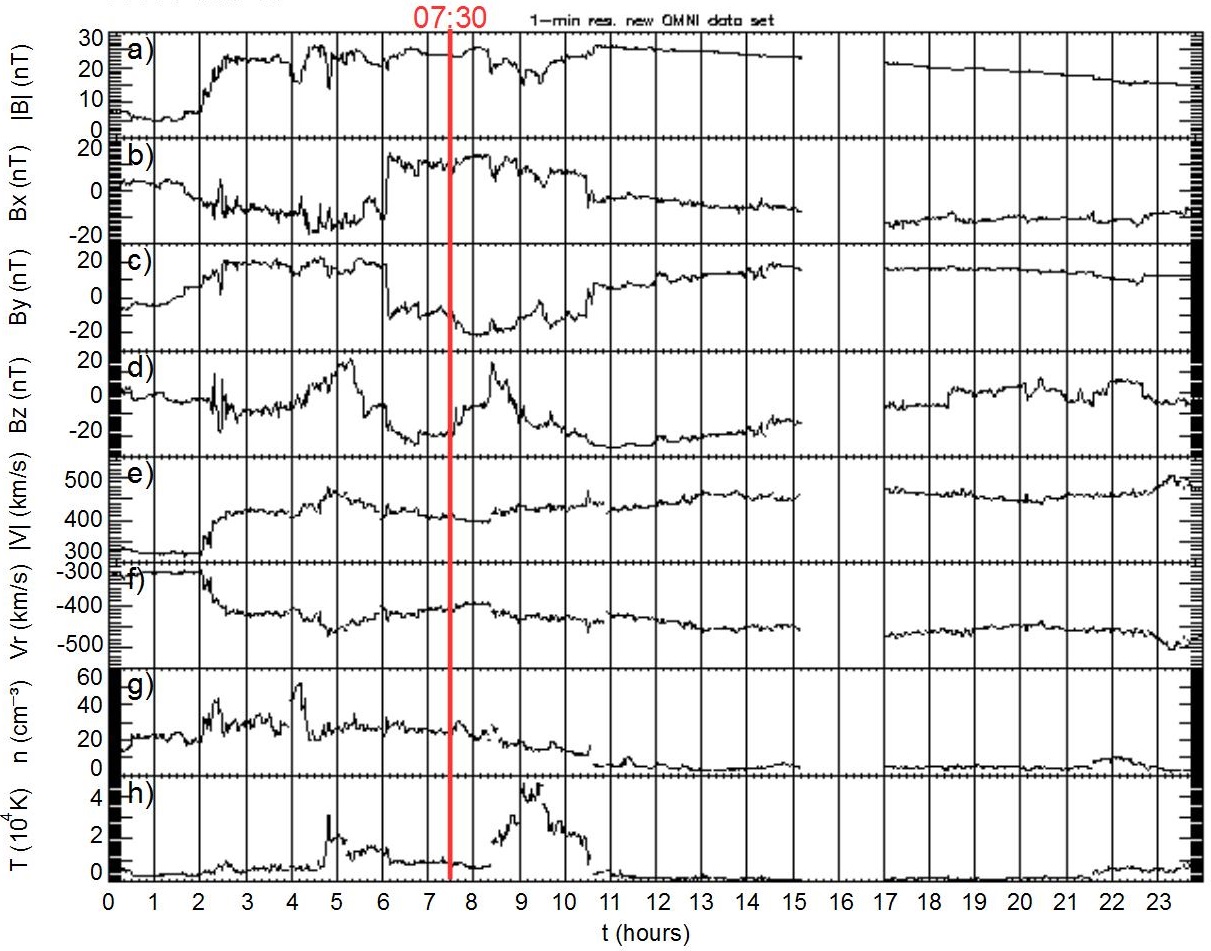}}
\caption{OMNIWeb data during the $15/05/1997$ ICME. (a) $|B|$ (nT), (b) $B_{x}$ (nT), (c) $B_{y}$ (nT), (d) $B_{z}$ (nT), (e) $|v|$ (km/s), (f) $v_{r}$ (km/s), (g) $n$ (cm$^{-3}$) and (h) T ($10^4$ K). The solid red line indicates the time frame selected as the simulation input.}
\label{x2}
\end{figure}

\begin{figure}[h]
\centering
\resizebox{\hsize}{!}{\includegraphics[width=\columnwidth]{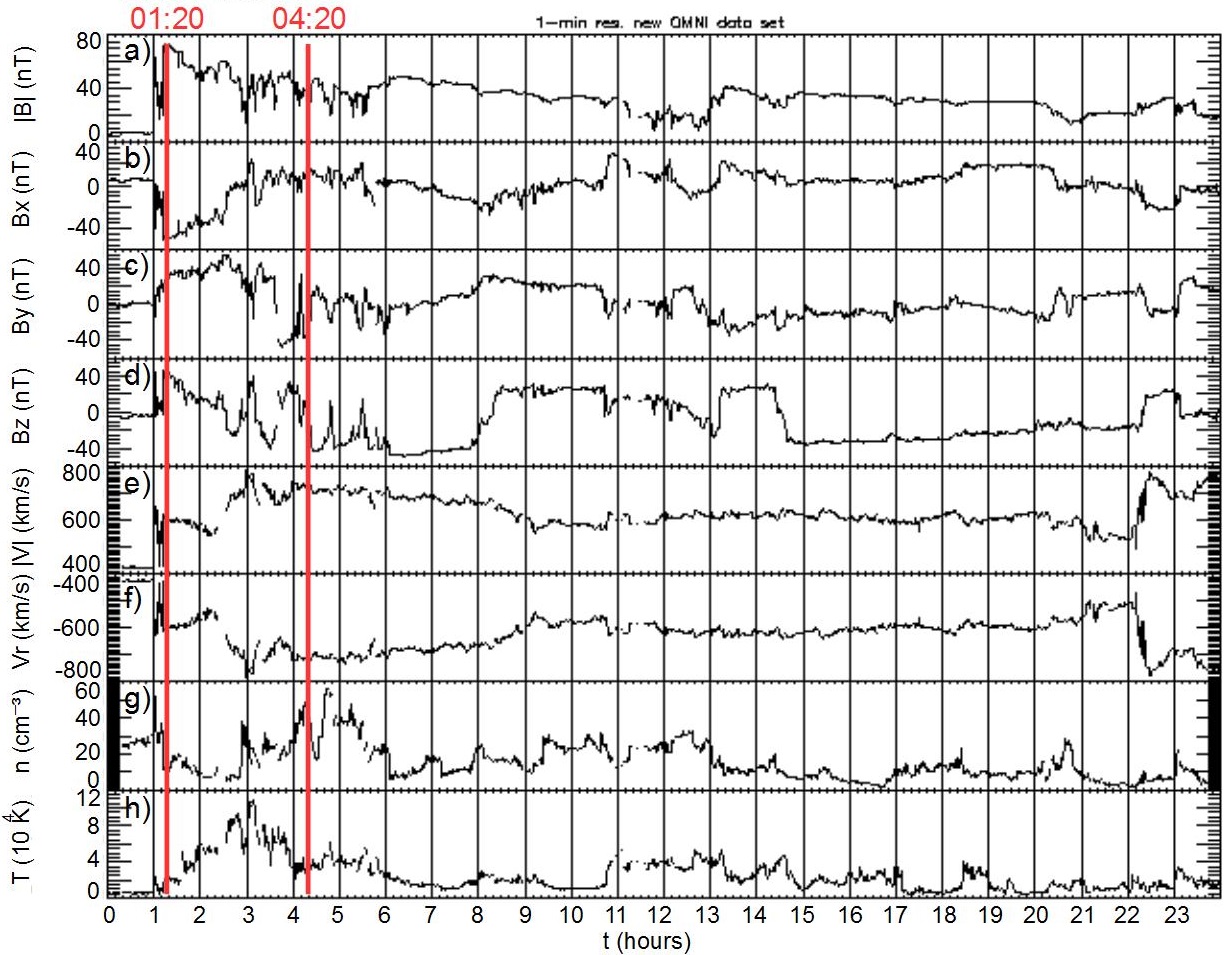}}
\caption{OMNIWeb data during the $31/03/2001$ ICME. (a) $|B|$ (nT), (b) $B_{x}$ (nT), (c) $B_{y}$ (nT), (d) $B_{z}$ (nT), (e) $|v|$ (km/s), (f) $v_{r}$ (km/s), (g) $n$ (cm$^{-3}$) and (h) T ($10^4$ K). The solid red line indicates the time frame selected as the simulation input.}
\label{x3}
\end{figure}

Figures \ref{x2} and \ref{x3} show two examples of the space weather condition obtained from OMNIWeb used as input of the simulations for the $15/05/1997$ and $31/03/2001$ ICMEs, respectively.

\section{Acronym list}

\begin{table}[h]
\centering
\begin{tabular}{c | c}
Acronym & Meaning \\ \hline
IMF & Interplanetary Magnetic Field  \\
SW & Solar Wind  \\
MHD & Magneto Hydro Dynamic \\
CME & Coronary Mass Ejection \\
ICME & Interplanetary Coronal Mass Ejection \\
$Dst$ & Disturbance Storm Time Index \\
BS & Bow Shock \\
DS & Day Side \\
NS & Night Side \\
OCB & Open-Close Boundary \\
\end{tabular}
\caption{Acronym list.}
\label{w}
\end{table}

\section{Parameter simulation list}

The SW and IMF parameters in the simulations included in the figures \ref{6}, \ref{10}, \ref{11}, \ref{12}, \ref{13}, \ref{14}, \ref{15} are (for Sun-Earth, Earth-Sun, Northward, Southward and Ecliptic ctr-clockwise IMF orientations):
\begin{table}[h]
\centering
\begin{tabular}{c c c c c}
$n$ (cm$^{-3}$) & $T$ ($10^{3}$ K) & $V_{r}$ (km/s) & $P_{d}$ (nPa) & $|B|_{sw}$ (nT) \\
\hline
$12$ & $180$ & $-350$ & $1.2$ & $10$ \\
$12$ & $180$ & $-350$ & $1.2$ & $20$ \\
$12$ & $180$ & $-350$ & $1.2$ & $30$ \\
$12$ & $180$ & $-350$ & $1.2$ & $40$ \\
$12$ & $180$ & $-350$ & $1.2$ & $50$ \\
$12$ & $180$ & $-350$ & $1.2$ & $60$ \\
$12$ & $180$ & $-350$ & $1.2$ & $70$ \\
$12$ & $180$ & $-350$ & $1.2$ & $80$ \\
$12$ & $180$ & $-350$ & $1.2$ & $90$ \\
$12$ & $180$ & $-350$ & $1.2$ & $100$ \\
$12$ & $180$ & $-350$ & $1.2$ & $125$ \\
$12$ & $180$ & $-350$ & $1.2$ & $150$ \\
$12$ & $180$ & $-350$ & $1.2$ & $175$ \\
$12$ & $180$ & $-350$ & $1.2$ & $200$ \\
$12$ & $180$ & $-350$ & $1.2$ & $225$ \\
$12$ & $180$ & $-350$ & $1.2$ & $250$ \\
\end{tabular}
\caption{Parameter list in the simulations included in figures \ref{6}, \ref{10}, \ref{11}, \ref{12}, \ref{13}, \ref{14}, \ref{15}.}
\label{x5}
\end{table}

The SW and IMF parameters in the simulations included in the figures \ref{8} and \ref{9} fixing the SW velocity and temperature are (for the Sun-Earth IMF orientation):
\begin{table}[h]
\centering
\begin{tabular}{c c c c c}
$n$ (cm$^{-3}$) & $T$ ($10^{3}$ K) & $V_{r}$ (km/s) & $P_{d}$ (nPa) & $|B|_{sw}$ (nT) \\
\hline
$6$ & $180$ & $-350$ & $0.6$ & $10$ \\
$6$ & $180$ & $-350$ & $0.6$ & $0$ \\
$12$ & $180$ & $-350$ & $1.2$ & $10$ \\
$12$ & $180$ & $-350$ & $1.2$ & $0$ \\
$18$ & $180$ & $-350$ & $1.8$ & $10$ \\
$24$ & $180$ & $-350$ & $2.4$ & $10$ \\
$24$ & $180$ & $-350$ & $2.4$ & $0$ \\
$30$ & $180$ & $-350$ & $3.1$ & $10$ \\
$36$ & $180$ & $-350$ & $3.7$ & $10$ \\
$36$ & $180$ & $-350$ & $3.7$ & $0$ \\
$42$ & $180$ & $-350$ & $4.3$ & $10$ \\
$48$ & $180$ & $-350$ & $4.9$ & $10$ \\
$48$ & $180$ & $-350$ & $4.9$ & $0$ \\
$54$ & $180$ & $-350$ & $5.5$ & $10$ \\
$60$ & $180$ & $-350$ & $6.1$ & $10$ \\
$60$ & $180$ & $-350$ & $6.1$ & $0$ \\
$72$ & $180$ & $-350$ & $7.4$ & $10$ \\
$84$ & $180$ & $-350$ & $8.6$ & $10$ \\
$96$ & $180$ & $-350$ & $9.8$ & $10$ \\
$108$ & $180$ & $-350$ & $11.0$ & $10$ \\
$120$ & $180$ & $-350$ & $12.3$ & $10$ \\
$135$ & $180$ & $-350$ & $13.8$ & $10$ \\
$150$ & $180$ & $-350$ & $15.3$ & $10$ \\
$165$ & $180$ & $-350$ & $16.9$ & $10$ \\
$180$ & $180$ & $-350$ & $18.4$ & $10$ \\
$195$ & $180$ & $-350$ & $19.9$ & $10$ \\
$210$ & $180$ & $-350$ & $21.5$ & $10$ \\
$240$ & $180$ & $-350$ & $24.5$ & $10$ \\
$270$ & $180$ & $-350$ & $27.6$ & $10$ \\
$300$ & $180$ & $-350$ & $30.7$ & $10$ \\
$330$ & $180$ & $-350$ & $33.7$ & $10$ \\
$360$ & $180$ & $-350$ & $36.8$ & $10$ \\
$400$ & $180$ & $-350$ & $41.0$ & $10$ \\
$450$ & $180$ & $-350$ & $46.1$ & $10$ \\
$500$ & $180$ & $-350$ & $51.2$ & $10$ \\
$550$ & $180$ & $-350$ & $56.3$ & $10$ \\
$600$ & $180$ & $-350$ & $61.5$ & $10$ \\
\end{tabular}
\caption{Parameter list in the simulations included in figures \ref{8} and \ref{9} fixing the SW velocity and temperature.}
\label{x6}
\end{table}

The SW and IMF parameters in the simulations included in the figures \ref{8} and \ref{9} fixing the SW density and temperature are (for the Sun-Earth IMF orientation):
\begin{table}[h]
\centering
\begin{tabular}{c c c c c}
$n$ (cm$^{-3}$) & $T$ ($10^{3}$ K) & $V_{r}$ (km/s) & $P_{d}$ (nPa) & $|B|_{sw}$ (nT) \\
\hline
$12$ & $180$ & $-100$ & $0.1$ & $10$ \\
$12$ & $180$ & $-100$ & $0.1$ & $0$ \\
$12$ & $180$ & $-150$ & $0.2$ & $10$ \\
$12$ & $180$ & $-200$ & $0.4$ & $10$ \\
$12$ & $180$ & $-200$ & $0.4$ & $0$ \\
$12$ & $180$ & $-250$ & $0.6$ & $10$ \\
$12$ & $180$ & $-300$ & $0.9$ & $10$ \\
$12$ & $180$ & $-300$ & $0.9$ & $0$ \\
$12$ & $180$ & $-350$ & $1.2$ & $10$ \\
$12$ & $180$ & $-400$ & $1.6$ & $10$ \\
$12$ & $180$ & $-400$ & $1.6$ & $0$ \\
$12$ & $180$ & $-450$ & $2.0$ & $10$ \\
$12$ & $180$ & $-500$ & $2.5$ & $10$ \\
$12$ & $180$ & $-500$ & $2.5$ & $0$ \\
$12$ & $180$ & $-550$ & $3.0$ & $10$ \\
$12$ & $180$ & $-600$ & $3.6$ & $10$ \\
$12$ & $180$ & $-600$ & $3.6$ & $0$ \\
$12$ & $180$ & $-650$ & $4.2$ & $10$ \\
$12$ & $180$ & $-700$ & $4.9$ & $10$ \\
$12$ & $180$ & $-750$ & $5.6$ & $10$ \\
$12$ & $180$ & $-800$ & $6.4$ & $10$ \\
$12$ & $180$ & $-850$ & $7.2$ & $10$ \\
$12$ & $180$ & $-900$ & $8.1$ & $10$ \\
$12$ & $180$ & $-950$ & $9.1$ & $10$ \\
$12$ & $180$ & $-1000$ & $10.0$ & $10$ \\
$12$ & $180$ & $-1100$ & $12.1$ & $10$ \\
$12$ & $180$ & $-1200$ & $14.4$ & $10$ \\
$12$ & $180$ & $-1300$ & $17.0$ & $10$ \\
$12$ & $180$ & $-1400$ & $19.7$ & $10$ \\
$12$ & $180$ & $-1500$ & $22.6$ & $10$ \\
$12$ & $180$ & $-1750$ & $30.7$ & $10$ \\
$12$ & $180$ & $-2000$ & $40.1$ & $10$ \\
$12$ & $180$ & $-2250$ & $50.8$ & $10$ \\
$12$ & $180$ & $-2500$ & $62.7$ & $10$ \\
$12$ & $180$ & $-2750$ & $75.9$ & $10$ \\
$12$ & $180$ & $-3000$ & $90.3$ & $10$ \\
$12$ & $180$ & $-3250$ & $106.0$ & $10$ \\
$12$ & $180$ & $-3500$ & $122.9$ & $10$ \\
$12$ & $180$ & $-3750$ & $141.1$ & $10$ \\
$12$ & $180$ & $-4000$ & $160.6$ & $10$ \\
\end{tabular}
\caption{Parameter list in the simulations included in figures \ref{8} and \ref{9} fixing the SW density and temperature.}
\label{x7}
\end{table}

The SW and IMF parameters in the simulations included in figure \ref{9} fixing the SW density and velocity are (for the Sun-Earth IMF orientation):
\begin{table}[h]
\centering
\begin{tabular}{c c c c c}
$n$ (cm$^{-3}$) & $T$ ($10^{3}$ K) & $V_{r}$ (km/s) & $P_{d}$ (nPa) & $|B|_{sw}$ (nT) \\
\hline
$12$ & $50$ & $-350$ & $1.2$ & $10$ \\
$12$ & $100$ & $-350$ & $1.2$ & $10$ \\
$12$ & $150$ & $-350$ & $1.2$ & $10$ \\
$12$ & $180$ & $-350$ & $1.2$ & $10$ \\
$12$ & $250$ & $-350$ & $1.2$ & $10$ \\
$12$ & $300$ & $-350$ & $1.2$ & $10$ \\
$12$ & $350$ & $-350$ & $1.2$ & $10$ \\
$12$ & $400$ & $-350$ & $1.2$ & $10$ \\
$12$ & $450$ & $-350$ & $1.2$ & $10$ \\
$12$ & $500$ & $-350$ & $1.2$ & $10$ \\
$12$ & $600$ & $-350$ & $1.2$ & $10$ \\
$12$ & $700$ & $-350$ & $1.2$ & $10$ \\
$12$ & $800$ & $-350$ & $1.2$ & $10$ \\
$12$ & $900$ & $-350$ & $1.2$ & $10$ \\
$12$ & $1000$ & $-350$ & $1.2$ & $10$ \\
\end{tabular}
\caption{Parameter list in the simulations included in figure \ref{9} fixing the SW velocity and velocity.}
\label{x8}
\end{table}

The SW and IMF parameters in the simulations included in the figures \ref{16} and \ref{17} fixing the SW density and temperature are (for Earth-Sun, Northward, Southward and Ecliptic ctr-clockwise IMF orientations):
\begin{table}[h]
\centering
\begin{tabular}{c c c c c}
$n$ (cm$^{-3}$) & $T$ ($10^{3}$ K) & $V_{r}$ (km/s) & $P_{d}$ (nPa) & $|B|_{sw}$ (nT) \\
\hline
$12$ & $180$ & $-350$ & $1.2$ & $[50-250]$ \\
$12$ & $180$ & $-385$ & $1.5$ & $[50-250]$ \\
$12$ & $180$ & $-545$ & $3.0$ & $[50-250]$ \\
$12$ & $180$ & $-670$ & $4.5$ & $[50-250]$ \\
$12$ & $180$ & $-775$ & $6.0$ & $[50-250]$ \\
$12$ & $180$ & $-1225$ & $15.0$ & $[50-250]$ \\
$12$ & $180$ & $-1730$ & $30.0$ & $[50-250]$ \\
$12$ & $180$ & $-2120$ & $45.0$ & $[50-250]$ \\
$12$ & $180$ & $-2445$ & $60.0$ & $[50-250]$ \\
$12$ & $180$ & $-2825$ & $80.0$ & $[50-250]$ \\
$12$ & $180$ & $-3160$ & $100$ & $[50-250]$ \\
\end{tabular}
\caption{Parameter list in the simulations included in figures \ref{16} and \ref{17} fixing the SW density and temperature. The $\Delta B$ between simulations is $50$ nT.}
\label{x9}
\end{table}
 
\end{appendix}


\bibliographystyle{aa}
\bibliography{References}

\end{document}